\documentclass[12pt]{article}
\pdfoutput=1
\usepackage{amsmath,amsfonts,amssymb,amsthm,amssymb,bbm,bm}
\usepackage{pdfsync}
\usepackage{cite}
\usepackage{graphicx,import}
\usepackage{enumerate}
\usepackage[latin1]{inputenc}
\usepackage{hyperref}
\usepackage{empheq}
\usepackage[all]{xy}
\usepackage{stmaryrd}
\usepackage{rotating}
\usepackage{color}  
\usepackage{slashed}
\usepackage{cancel}
\usepackage{array, makecell} %
\usepackage{comment}
\usepackage{tikz-cd}
\usepackage{hhline}
\usepackage{comment}
\usepackage{cancel}
\usepackage{subcaption}
\usepackage[font=footnotesize]{caption}
\makeatletter

\DeclareFontFamily{U}{matha}{\hyphenchar\font45}
\DeclareFontShape{U}{matha}{m}{n}{
      <5> <6> <7> <8> <9> <10> gen * matha
      <10.95> matha10 <12> <14.4> <17.28> <20.74> <24.88> matha12
      }{}
\DeclareSymbolFont{matha}{U}{matha}{m}{n}
\DeclareMathSymbol{\oright}       {2}{matha}{"69}

\usetikzlibrary{calc}
\usetikzlibrary{positioning}
\usetikzlibrary{arrows,backgrounds}
\usetikzlibrary{patterns}
\usetikzlibrary{decorations.pathreplacing,decorations.markings}
\usetikzlibrary{decorations.pathmorphing}
\usetikzlibrary{shapes}

\pgfdeclarelayer{background}
\pgfdeclarelayer{foreground}
\pgfsetlayers{background,foreground}

\usetikzlibrary{backgrounds}
\usetikzlibrary{patterns}
\usetikzlibrary{arrows.meta}
\usetikzlibrary{shapes}

\newcommand{\doublehat}[1]{%
\begingroup%
  \let\macc@kerna\z@%
  \let\macc@kernb\z@%
  \let\macc@nucleus\@empty%
  \hat{\raisebox{.55ex}{\vphantom{\ensuremath{#1}}}\smash{\hat{#1}}}%
\endgroup%
}

\newcommand{\p}{\partial}

\newcommand{\bit}{\begin{itemize}}
\newcommand{\eit}{\end{itemize}}
\newcommand{\bd}{\begin{description}}
\newcommand{\ed}{\end{description}}

\newcommand{\bc}{\begin{center}}
\newcommand{\ec}{\end{center}}

\newcommand{\C}{{\mathbb C}}
\newcommand{\N}{{\mathbb N}}
\newcommand{\R}{{\mathbb R}}
\newcommand{\Z}{{\mathbb Z}}

\newcommand{\cM}{{\cal M}}
\newcommand{\cP}{{\cal P}}
\newcommand{\cJ}{{\cal J}}
\newcommand{\cN}{{\cal N}}
\newcommand{\cT}{{\cal T}}

\newcommand{\cL}{{\mathcal L}}

\newcommand{\cQ}{{\mathcal{Q}}}

\def\be#1\ee{\begin{align}#1\end{align}}
\newcommand{\bea}{\begin{eqnarray}}
\newcommand{\eea}{\end{eqnarray}}
\newcommand{\bs}{\begin{subequations}}
\newcommand{\es}{\end{subequations}}

\newcommand{\la}{\label}

\newcommand{\f}{\frac}

\newcommand{\bmm}{\bm{m}}
\newcommand{\bz}{{\bar{z}}}

\def\p{\partial}

\def\N{N}

\def\d{\delta}

\def\rd{\mathrm{d}}
\def\pa{\partial }

\def\k{{\kappa^2} }

\newcommand{\tcM}{{\tilde{\mathcal M}}}

\newcommand{\scri}{\cal I}

\begin{document}

\title{\Large{\bf
Higher spin dynamics in gravity and $w_{1 + \infty}$ celestial symmetries}}

\author{ Laurent Freidel$^1$\thanks{lfreidel@perimeterinstitute.ca} , Daniele Pranzetti$^{1,2}\thanks{dpranzetti@perimeterinstitute.ca}$ ,
Ana-Maria Raclariu$^1$\thanks{araclariu@perimeterinstitute.ca}
}
\date{\small{\textit{
$^1$Perimeter Institute for Theoretical Physics,\\ 31 Caroline Street North, Waterloo, Ontario, Canada N2L 2Y5\\ \smallskip
$^2$ Universit\`a degli Studi di Udine, via Palladio 8,  I-33100 Udine, Italy
}}}

\maketitle

\begin{abstract}
In this paper we extract from a large-$r$ expansion of the vacuum Einstein's equations a dynamical system governing the time evolution of  an infinity of higher-spin charges. Upon integration, we evaluate the canonical action of these charges on the gravity phase space.
The truncation of this action to quadratic order and the associated charge conservation laws yield an infinite tower of soft theorems. We show that the canonical action of the higher spin charges on gravitons in a conformal primary basis, as well as  conformally soft  gravitons reproduces the higher spin celestial symmetries derived from the operator product expansion. Finally, we give direct evidence that these charges form a canonical representation of a $w_{1+\infty}$ loop algebra on the gravitational phase space. 
\end{abstract}

\newpage
\tableofcontents

\section{Introduction}

What are the symmetries of gravitational theories? Are these symmetries enough to determine gravitational dynamics? These questions have proven central to the quest of  uncovering the nature of quantum gravity and revealed new connections among different areas of physics. 

In the presence of gravitational radiation, the asymptotic symmetry group of four-dimensional asymptotically flat spacetimes (AFS) is necessarily infinite dimensional \cite{Bondi:1960jsa, BMS, Sachs62}. However, determining the set of boundary conditions and resulting asymptotic symmetries that accommodate all physical gravitational phenomena is a challenging task. In recent years, an imprint of asymptotic symmetries in the gravitational S-matrix was discovered: the soft graviton theorem \cite{Weinberg:1965nx} was shown to arise from conservation of BMS supertranslation charges \cite{Strominger:2013jfa, He:2014laa}. A wealth of surprising connections followed, from new soft graviton theorems \cite{Cachazo:2014fwa, White:2014qia}, to new asymptotic symmetries \cite{Barnich:2009se, Barnich:2010eb, Barnich:2011mi, Barnich:2013axa, Kapec:2014opa, Campiglia:2014yka, Campiglia:2015yka, Flanagan:2015pxa, Compere:2018ylh, Campiglia:2020qvc, Freidel:2021yqe} and memory effects \cite{Pasterski:2015tva, Nichols:2017rqr, Nichols:2018qac}. The latter 
turned the art of choosing the ``right'' boundary conditions into more of a science by providing physical criteria to 
single out asymptotic diffeomorphisms that should be promoted to symmetries. More generally, the link between the S-matrix program and  celestial holography -- a recently proposed holographic description for gravity in AFS -- \cite{Pasterski:2016qvg, Pasterski:2017kqt} (see \cite{Raclariu:2021zjz, Pasterski:2021rjz, Pasterski:2021raf} for reviews) may lead to further constraints, in particular by revealing and exploiting new symmetries \cite{Banerjee:2018fgd, Banerjee:2018gce, Donnay:2018neh, Stieberger:2018onx, Pate:2019lpp, Pate:2019mfs, Banerjee:2020zlg, Banerjee:2021cly, Banerjee:2021dlm, Guevara:2021abz, Strominger:2021lvk}.

On the asymptotic symmetry front, these developments emphasized the importance of properly accounting for boundary degrees of freedom \cite{Ashtekar:1981bq,Dray:1984rfa,Ashtekar:1990gc}  and led to a revision of the allowed boundary conditions and resulting symmetry algebras \cite{Barnich:2011ty, Barnich:2013axa, Barnich:2016lyg, Compere:2018ylh, Freidel:2021yqe}. At the same time a considerable amount of progress was achieved at finite distance where, on the one hand, the central concept of corner symmetry revealed new types of infinite dimensional symmetries playing a key role in the decomposition of gravitational systems into subsystems \cite{Freidel:2015gpa,Donnelly:2016auv, Freidel:2020xyx, Freidel:2020svx, Freidel:2020ayo, Freidel:2021cjp, Ciambelli:2021vnn, Freidel:2021dxw, Ciambelli:2021nmv, Chandrasekaran:2021vyu}. 
 On the other hand,  new approaches in analyzing the gravitational phase space of black hole horizons and  null surfaces \cite{Hopfmuller:2016scf,Hopfmuller:2018fni, Donnay:2015abr,Donnay:2016ejv,Donnay:2019jiz, Chandrasekaran:2018aop, Adami:2020amw, Grumiller:2019ygj, Grumiller:2019fmp, Adami:2021nnf} have been proposed. Finally, a new understanding of the gravitational renormalization procedure connecting finite to asymptotic surfaces has been achieved \cite{Compere:2008us, Compere:2018ylh, Freidel:2021yqe, Chandrasekaran:2021vyu }. On the celestial side, some of the highlights include the reformulation of scattering amplitudes into a basis of asymptotic boost eigenstates \cite{Pasterski:2016qvg, Pasterski:2017kqt,Pasterski:2017ylz,Arkani-Hamed:2020gyp, Brandhuber:2021nez}, an ever-growing catalogue of celestial symmetries \cite{Stieberger:2018onx,Fotopoulos:2019tpe, Fotopoulos:2019vac, Pate:2019mfs,Adamo:2019ipt, Puhm:2019zbl, Fan:2020xjj,Donnay:2020guq,Fotopoulos:2020bqj,Pano:2021ewd,Pasterski:2021fjn,Pasterski:2021dqe} and their associated constraints \cite{Law:2019glh,Pate:2019lpp,Banerjee:2020zlg,Banerjee:2021cly,Banerjee:2021dlm}, as well as a framework amenable to the use of standard conformal field theory (CFT) methods \cite{Nandan:2019jas,Law:2020xcf,Fan:2021isc,Atanasov:2021cje,Fan:2021pbp} for gravity in AFS. Intriguingly, a $w_{1 + \infty}$ structure \cite{Bakas:1989um, Pope:1991ig, Shen:1992dd} was recently encountered in the algebra of the infinite tower of conformally soft graviton symmetries \cite{Guevara:2021abz, Strominger:2021lvk}.  The origin of this symmetry was explained in the context of the ambitwistor string \cite{Adamo:2021lrv, Adamo:2021zpw} and self-dual gravity \cite{Ball:2021tmb}. Nevertheless, the original derivation of the $w_{1 + \infty}$ structure is agnostic to the type of gravitational theory and should universally govern gravitational scattering at tree level.   If true this suggests that it should also constrain the classical  gravitational dynamics. It  seems therefore imperative to look for such higher-spin symmetries in Einstein gravity and to understand their spacetime interpretation. 

As shown in \cite{Guevara:2021abz}, an entire tower of soft symmetries is generated as soon as the generalized BMS symmetries \cite{Barnich:2009se,Fotopoulos:2019vac} are supplemented by the sub-subleading soft graviton symmetry. In \cite{Freidel:2021dfs} we demonstrated that the sub-subleading soft graviton theorem arises as a consequence of the conservation of a spin-2 charge whose evolution is dictated by one of the leading order asymptotic Einstein's equations in a large-$r$ expansion. This is in close analogy with the leading and subleading soft graviton theorems which were found to arise from conservation of Bondi mass and angular momentum aspects \cite{Strominger:2013jfa, He:2014laa, Kapec:2014opa} associated with the remaining components of the asymptotic equations of motion at the same order. Equivalently, all universal soft theorems can be understood as resulting from matching conditions on the leading asymptotic components of the Newman--Penrose scalars \cite{NP62, Newman:1962cia, Adamo:2009vu}. 

The main goal of this work is to extend the analysis of \cite{Freidel:2021dfs} to a tower of higher-spin charges {obeying} the following recursion relations
\be\la{dotQss-intro}
\dot{\cQ}_{s} = D \cQ_{s-1} + \frac{(1+s)}{2} C \cQ_{s-2}\,, \quad s \geq -1,~~ s \in \mathbb{Z}.
\ee
Here $C$ is related to the asymptotic shear, while for $s = 0, 1, 2,$ $\cQ_s$ correspond to the Bondi mass and angular momentum aspects, and the spin-2 tensor respectively \cite{Freidel:2021qpz}.
Using results of \cite{Newman:1968uj}, we verify explicitly that $\cQ_3$ appears as a subleading term in the  asymptotic expansion of the Weyl scalar $\Psi_0$ that captures the incoming radiation. For $s\geq 4$ we conjecture that the simple evolution \eqref{dotQss-intro}  corresponds to a truncation of the evolution equations for all subleading terms in a large-$r$ asymptotic expansion of $\Psi_0$. 
In the linearized theory, the recursion relations \eqref{dotQss-intro}  imply an infinite set of conserved quantities associated with the presence of incoming radiation  first pointed out in
\cite{Newman:1968uj}.
We provide further evidence for the physical relevance of \eqref{dotQss-intro}   and the role of $\Psi_0$ in capturing the correct gravitational dynamics  by demonstrating that \eqref{dotQss-intro}, truncated to quadratic order in the fields, is precisely equivalent to the tower of soft higher-spin symmetries found in \cite{Guevara:2021abz, Strominger:2021lvk} by CFT methods.

Motivated by this connection, we embark in a canonical analysis to derive the charge bracket of a properly renormalized version of the higher spin charges, denoted $Q_s(\tau)$ with $\tau$ a smearing transformation parameter on the celestial sphere. We restrict our analysis to the linear (in the radiative data) contribution to the bracket and show that the loop algebra $L w_{1+\infty}$  is indeed realized within the phase space of gravity. Explicitly, we find the bracket \footnote{The superscript $1$ denotes the truncation to linear order, while $D$ is the 2d covariant derivative on the celestial sphere.}
\be
\{Q_s(\tau),  Q_{s'}(\tau')\}^1=
 Q^1_{s'+s-1}\left[ \left(s'+1\right) \tau' D\tau
- (s+1) \tau D\tau' \right]\,.\la{WTF-intro}
\ee

This paper is organized as follows. In Section \ref{sec:EOM} we review the derivation of \eqref{dotQss-intro} for $s \in \{-1, 0, 1, 2\}$ from symmetry arguments \cite{Freidel:2021qpz}, namely relying on the re-organization of asymptotic data  in terms of primaries with respect to the homogeneous subgroup of the Weyl-BMS  group   \cite{Freidel:2021yqe}, as well as the boundary conditions necessary in order to establish the leading, subleading and sub-subleading soft theorems.  In Section \ref{hss}, we discuss the generalization of these conditions to higher spins. In Section \ref{subs:hssg} we identify the $s = 3$ charge in the gravitational asymptotic phase space as the next-to-leading component in a large-$r$ expansion of $\Psi_0$ and recast the corresponding evolution equation identified in \cite{Barnich:2019vzx} into \eqref{dotQss-intro}. The same recursion relation is solved for arbitrary $s$ in terms of the news at linear and quadratic orders in the fields in Section \ref{ss:hssa}, and the quadratic action on the shear is computed. 
In Section \ref{sec:pseudo} we use this action to derive the pseudo-vector fields associated to the higher spin charge transformations, generalizing the result of \cite{Freidel:2021dfs} to $s\geq 3$.

In Section \ref{sec:soft-th} we show that the conservation of these higher-spin charges truncated to quadratic order is equivalent to the infinite tower of conformally soft theorems discussed in \cite{Guevara:2021abz, Strominger:2021lvk}.
In Section \ref{ss:celestial} we prove that this action matches exactly  the action
of the infinity of celestial soft symmetries implied by the celestial operator product expansion (OPE) block \cite{Guevara:2021abz}.
In Section \ref{sec:w-structure} we review the celestial diamond structure pointed out in \cite{Pasterski:2021fjn, Pasterski:2021dqe, Guevara:2021abz}, extend this structure to a general (sub)$^s$-leading soft graviton and identify its dual as the order $s$ sub-leading component of $\Psi_0$.
In Section \ref{sec:wcurr} we clarify the definition of the light transform of the soft graviton, as well as its relation to the $w$-currents identified with the generators of the wedge subalgebra of $w_{1 + \infty}$ symmetry in \cite{Himwich:2021dau, Jiang:2021ovh} and the canonical soft charges.
The OPEs of the latter two quantities are compared
in Section \ref{sec:recursion}, revealing an intriguing connection between the two sets of global and canonical charges. 
Finally, in Section \ref{sec:CB} the bracket \eqref{WTF-intro} is derived.
Some technical details are collected in Appendices \ref{App:SGbra}, \ref{App:pseudo},\ref{App:barC}, \ref{app:CB}, \ref{App:Norm} and \ref{App:wcurr}.


\section{Preliminaries}\la{sec:EOM}

In \cite{Freidel:2021qpz} it was shown that the asymptotic Einstein's equations can be recovered by identifying the Weyl scalars at null infinity. Moreover, a well-defined notion of non-radiative corner phase space was proposed. That analysis relies on the observation that asymptotic charges are primary fields with respect to the homogeneous subgroup ${H}_S:=(\mathrm{Diff}(S)\ltimes \mathrm{Weyl})$ of the Weyl-BMS (BMSW) group \cite{Freidel:2021yqe}. BMSW is an extension of the original BMS group \cite{Bondi:1960jsa, BMS, Sachs62} of gravitational symmetries of null infinity by arbitrary diffeomorphisms on the celestial 2-sphere $S$. ${H}_S$ is generated by vector fields $Y^A(\sigma^A)$ and local Weyl rescalings $W(\sigma^A)$ on the sphere, while BMSW$= H_S\ltimes \R^S$ also includes supertranslations parametrized by a function $T(\sigma^A).$
 We work in Bondi coordinates where $\sigma^A$ are coordinates on $S$ and $u$ is the retarded time along $\scri^+$. 
 
\subsection{Asymptotic phase space}
\la{sec:AsymPS}

  For a given cut $u=0$ of $\scri^+$, primary fields $O_{(\Delta,s)}$ are defined by their transformation law with respect to $H_S,$
\be\la{eq:prim-trans} 
\delta_{(Y,W)} O_{(\Delta,s)}
= (\cL_Y + (\Delta-s) W ) O_{(\Delta,s)},
\ee
where $\cL_Y$ is the Lie derivative along $Y$. They are labelled by their spin $s$ and conformal dimension $\Delta$.  
Some of the primary fields represent radiative degrees of freedom, namely the shear $C_{AB}$ and 
the shifted news tensor $\hat{N}^{AB}$ defined by
\be\la{hN}
\hat{N}^{AB}:= N^{AB}- \tau^{AB}\,.
\ee 
Here $N^{AB}:=\dot{C}^{AB}$ and $\tau^{AB}$  is the symmetric traceless Geroch tensor \cite{Geroch:1977jn} defined by the condition $D_A \tau^{AB} +\f12 D^B R =0$, where $D_A$ is the covariant derivative associated with the 2-sphere metric and $R$ is the 2d Ricci scalar. The time derivative of the news is also a primary field which we denote by $\cN^{AB}:= \p_u \hat N^{AB}$. 

Other primary fields correspond to  asymptotic charges and label the non-radiative corner phase space when the no-radiation condition $\cN^{AB}=0$ is imposed \cite{Freidel:2021qpz}. They include the energy current $\cJ^A$, the covariant mass $\cM$, the covariant dual mass $\tcM$, the covariant momentum $\cP_A$ and the spin-$2$ tensor $\cT_{AB}$. 
 The spinning primaries can be traded for helicity- or spin-weighted scalars by contraction with frame fields, namely
\begin{eqnarray} \la{scalars}
&C:= C_{AB} m^A m^B\,,\quad
\hat N:=\hat N^{AB}\bar{m}_A\bar{m}_B\,,\quad
\cN:=\cN^{AB}\bar{m}_A\bar{m}_B\,,\quad
\cJ:= \cJ^A\bar{m}_A\,,\cr
&\cM_{\C} := \cM+i\tcM\,,\quad 
\cP:= \cP_A {m}^A\,,\quad
\cT:= \cT_{AB} {m}^A{m}^B\,.
\end{eqnarray}
We have introduced a holomorphic frame $m= m^A\pa_A$ with coframe $\bmm = m_A \rd \sigma^A$ normalized such that $m^A\bar{m}_A=1$. Contractions with $m_A$ and  $\bar{m}_A$ contribute helicity $+1$ and $-1$ respectively.\footnote{Both $m_A$ and $\bar{m}_A$ have dimension-spin $(\Delta, s) = (0,1)$, but opposite helicity. Assigning helicity $+1$ to  $m_A$ is conventional.} We use the same label $s$ to denote the helicity of a spin $s$ primary upon contraction with frame fields, namely
\be \label{spins}
O_s = O_{A_1\cdots A_s} m^{A_1}\cdots m^{A_s},
\qquad 
O_{-s} = O^{A_1\cdots A_s} \bar{m}_{A_1}\cdots  \bar{m}_{A_s}.
\ee 
We also define $D O_s = m^A m^{A_1} \cdots m^{A_s} D_A O_{A_1 \cdots A_s}$, where $D_A$ is the covariant derivative on the sphere.
One needs to recall that $(D, \pa_u)$ are operators that  respectively raise the  dimension/helicity by $(1,1)$ and $(1,0)$.

In the presence of radiation, the helicity  scalars associated to the symmetry charges can then be shown to obey the following evolution equations
\cite{Barnich:2011mi,Nichols:2018qac, Freidel:2021qpz, Freidel:2021dfs}
\begin{subequations}\label{eom}
\be
\dot{\cJ} &=\tfrac12  D \cN\,,\la{cJ}\\
\dot{\cM}_{ \C}&= D \cJ + \tfrac14 C \cN\,, \la{cM}\\
\dot{\cP}&= D\cM_{ \C} + C \cJ\,, \la{cP}\\
\dot{ \cT}&= D \cP + \tfrac32 C\cM_{ \C}\,,\la{cT}
\ee
\end{subequations}
and their complex conjugates. It will be convenient to relabel the gravitational data according to helicity and define
 \be\la{Qscalar}
 \cQ_{-2}:=\frac{\cN}{2},\qquad  \cQ_{-1}:= \cJ,\qquad \cQ_0:=\cM_\C,\qquad \cQ_1:=\cP,\qquad \cQ_2:=\cT.
 \ee
Then \eqref{cJ}-\eqref{cT} simply become
\be\la{dotQss}
\dot{\cQ}_{s} = D \cQ_{s-1} + \frac{(1+s)}{2} C \cQ_{s-2}\,,
\ee
for respectively $s=-1,0,1,2 $. 
The primary scalars \eqref{Qscalar} can be identified with
the leading terms in an asymptotic expansion of the 5 Weyl scalars (see \cite{Freidel:2021qpz} and Section \ref{subs:hssg} below). Note that the dimension/helicity of the charges are $(\Delta, J)=(3,s)$. We   summarize all the helicity-weighted scalars in Table \ref{tab:semi-primaries}.
\begin{table}[h] 
\centering
\begin{tabular}{|c|c|c| c| c | c| c| c|c|c|c|} 
 \hline
 Primary Scalars & $C$ & $\hat N$ &$\cN$ & $\cJ$ & $\cM_\C$ &  ${\cP}$ & ${\cT}$ & $\cQ_s$\\ 
 \hline 
Dimension-Helicity $(\Delta, J)$   &  (1,2)  & (2,-2)&  (3,-2)  & (3,-1) & (3,0) & (3,1) &(3,2)& (3,$s$)  \\
 \hline
 \end{tabular}
 \caption{ Conformal dimension and helicity of primary scalars.}
 \la{tab:semi-primaries}
 \end{table}

The asymptotic equations imply that the charges are functionals of the shear and the shifted news \eqref{hN}, which represent a pair of conjugate variables on  $\scri^+$. Their bracket takes the form \cite{He:2014laa, Ashtekar:1978zz, Ashtekar:1981sf, Ashtekar:2018lor}
\be \label{NCbra}
\{ \hat N(u,z), C(u', z') \} &=  \f\k2  \delta(u-u') \delta(z,z')\,,
\ee
with $\kappa=\sqrt{32\pi G}$.

\subsection{Mode expansions}\la{sec:exp}

At $\mathcal{I}^+$, the shear $C$ and its conjugate the (shifted) news ${\hat N}:=\pa_u \bar{C}$ admit the 
mode expansions \cite{Strominger:2013jfa}\footnote{ Polarisation factors are included in  our definition of $C= C_{AB} m^A m^B$.} 
\be 
\la{mode-exp}
C(u, \hat{x}) &= \frac{i\kappa}{8\pi^2}  \int_0^{\infty} d\omega \left[a_-^{\rm out\dagger}(\omega \hat{x}) e^{i\omega u} - a_{+}^{\rm out}(\omega\hat{x}) e^{-i\omega u} \right]\,,\\
\la{Nexp}
\hat N(u, \hat{x}) &= -\frac{\kappa}{8\pi^2} \int_0^{\infty} d\omega \omega \left[a_+^{\rm out\dagger}(\omega \hat{x}) e^{i\omega u} + a_{-}^{\rm out}(\omega\hat{x}) e^{-i\omega u} \right]\,,
\ee
for outgoing gravitons of momenta $q=\omega \hat x$.
At the quantum level, the bracket \eqref{NCbra} is then replaced by the commutator
\be \label{NCC}
[ \hat N(u,z), C(u', z') ] &= {-i} \f\k2  \delta(u-u') \delta(z,z')\,,
\ee
which implies the standard commutation relations for the oscillators
\be \label{comm}
 [a_-(\omega \hat{x}), a_-^\dagger(\omega' \hat{x}')] &=(2\pi)^3  \frac{2}{\omega}  \delta(\omega-\omega')\delta(z,z').
\ee

 In \cite{Freidel:2021dfs} we used  the commutator \eqref{NCC}, together with the boundary conditions  at  the future of $\scri^+$
\be\la{alpha}
 \cQ_s = O(u^{1+s-\alpha})\,,~~ C= O(u^{-\alpha})\,, \quad {\rm with}\quad  \alpha>3 \quad {\rm when}\quad u\to +\infty\,,
 \ee
 to show how the leading, subleading and, in particular, the sub-subleading soft graviton theorems \cite{Weinberg:1965nx, Cachazo:2014fwa} follow directly from the charge evolution equations 
 \eqref{dotQss}. To this end, a charge renormalization procedure was necessary and antipodal matching conditions \cite{Strominger:2013jfa}, as well as crossing symmetry at the S-matrix level were used. In the sub-subleading case, we found that the evolution equation for $\cT$ yields a tree-level collinear contribution to the soft graviton factor subleading in $\kappa$ which corrects the original analysis of \cite{Cachazo:2014fwa, Broedel:2014fsa, Bern:2014vva}.

 The main goal of this work is to show that the extension of \eqref{dotQss} beyond $s = 2$ encodes, after truncation to quadratic order, the tower of soft graviton symmetries \cite{Guevara:2021abz, Strominger:2021lvk} uncovered by completely different methods.

\section{Higher spin symmetry}
\label{hss}

As reviewed in the previous section, the asymptotic Einstein's equations at leading order in a large-$r$ expansion can be recast into the form
\eqref{dotQss}
for $s=-1,0,1,2 $ \cite{Freidel:2021dfs}.  One of the main results of our paper is that the extension of \eqref{dotQss}
to all $s \geq 3$
is responsible for the infinite tower of soft symmetries studied in \cite{Guevara:2021abz, Strominger:2021lvk}. 
In this section we use the results of \cite{Newman:1968uj} to explicitly verify this proposal for $s=3$ and argue that these equations appear as a truncation of the asymptotic Einstein's equations at subleading orders in a large-$r$ expansion. Moreover, we compute the action of the linear and     quadratic components of $\cQ_s$ on $C$ for all integer $s \geq 3$. In the next section we provide evidence for \eqref{dotQss} from celestial holography for all other higher spins, $s > 3$. In particular, we show that the truncation of \eqref{dotQss} to quadratic order in the fields implies the $w_{1+\infty}$ algebra structure revealed by \cite{Guevara:2021abz, Strominger:2021lvk}.

As a preliminary step to our analysis, we note that in order to integrate \eqref{dotQss} for {all} higher spins $s$, we need to assume that 
\be\la{Nbc}
\hat{N}=O(|u|^{-1-s-\epsilon})\,,\qquad {\rm with }~\epsilon>0\,,
\ee
and that  the geometry reverts to the vacuum at late retarded times, namely 
\be\la{Qbound}
\lim_{u\to +\infty}\cQ_s = 0\,.
\ee
This allows us to integrate \eqref{dotQss} resulting in the following  recursion relations among the higher spin charges
\be\la{highQ}
\cQ_{s} &= D\pa_u^{-1} (\cQ_{s-1}) + \frac{(s+1)}{2} \pa_u^{-1}(C \cQ_{s-2})\,.
\ee
We introduced the symbolic notation 
\be
 (\pa_u^{-n} \cQ)(u) := \int_{+\infty}^{u}\rd u_1 \int_{+\infty}^{u_1} \rd u_2 \cdots \int_{+\infty}^{u_{n-1}} \rd u_n\, \cQ(u_n) \,,\label{iiint}
\ee
where the order of integral labels is tailored to the choice of boundary conditions  \eqref{Qbound}. Since $\pa_u^{-1} D$ shift the dimension/helicity by $(0,1)$, $\cQ_s$ has $(\Delta,J)=(3,s)$. All higher spin charges have the same dimension $\Delta = 3$.

The recursion relation \eqref{highQ} can be solved by expanding each charge according to the number of oscillator fields it contains, namely
\be\cQ_s =\sum_{k=1}^{\max[2,s+1]} \cQ_s^{k}.
\la{charge-kexp}
\ee
In particular, $\cQ^{1}_s$ is the soft charge (linear in oscillators), while $\cQ^{2}_s$ is the hard charge including the quadratic (or free) contributions to the charge. $\cQ_s^{k}$ for $k\geq 3$ include collinear contributions of order $k$. Such contributions are present due to the non-linearity of Einstein's equations and are suppressed by powers of $G_N$. Nonlinear contributions to the spin $s$ charge have degree at most $s+1$ for $s\geq 1$.

\subsection{Higher spin symmetry from gravity}
\label{subs:hssg}
In this section we identify the spin-$3$ components from the gravity phase space. To this end, we first recall how  the covariant aspects \eqref{Qscalar} appear in the asymptotic expansion of the Weyl tensor. Consider
an asymptotically flat metric in the Bondi gauge \footnote{The explicit large-$r$ expansion of the metric coefficients is not crucial for the rest of our analysis and we refer the reader to \cite{Freidel:2021dfs} for it.}
\be \label{ds}
\rd s^2 = -2 e^{2\beta} \rd u (\rd r + \Phi \rd u) + r^2 \gamma_{AB} \left(d\sigma^A - \frac{\Upsilon^A}{r^2} \rd u\right)\left(d\sigma^B - \frac{\Upsilon^B}{r^2}\rd u \right)\,,
\ee
and introduce the null frame fields
\be
\ell = \pa_r, 
\qquad
n= e^{-2\beta}(\pa_u -\Phi \pa_r +r^{-2}\Upsilon^A\pa_A)\,.
\ee
The Weyl scalars
 are defined by
\be 
\Psi_0= -C_{\ell m \ell m}\,, \quad
\Psi_1= -C_{\ell n \ell m}\,, \quad
\Psi_2= -C_{\ell m \bar{m}  n} \,,
\quad\Psi_3= -C_{n \bar{m} n\ell }\,, \quad
\Psi_4= -C_{n \bar{m} n \bar{m} }\,,
\ee 
where $C_{abcd}$ is the Weyl tensor, $C_{\ell  m  \bar{m} n} :=C_{abcd}\ell^a  m^b \bar{m}^c n^d$ and similarly for the other contractions.
$\Psi_4$ represents the outgoing radiation at $\scri^+$ while $\Psi_0$ encodes the incoming radiation.

Their asymptotic expansions take the form
\be
\Psi_{2-s} =\frac{1}{r^{3+s}} \cQ_s - \frac{1}{r^{4+s}} \bar{D} \cQ_{s+1} + \cdots .
\ee 
 We see that the spin-$3$ charge $\cQ_3$ appears in the next-to-leading order expansion of $\Psi_0$. To confirm this we use the result of
  \cite{Newman:1968uj} (see also \cite{Barnich:2019vzx}), where it was shown that 
that this coefficient satisfies the evolution equation
\be 
\dot{\cQ}_3 = D \cQ_{2} + 2 C \cQ_{1},
\ee 
in agreement with \eqref{dotQss}.
We expect the higher spin charges to arise in the expansion of $\Psi_0 =\frac1{r^5}\sum_{n=0}^\infty{r^{-n}}\Psi_0^{(n)}$ in the form
\be 
 \Psi_{0} =\frac{1}{r^{5}} \cQ_2 - \frac{1}{r^{6}} \bar{D} \cQ_{3} +\sum_{s\geq 4}\frac1{r^{3+s}} \frac{(-1)^s}{ (s-2)! }\left( \bar{D}^{s-2} \cQ_s +  \cdots \right)\,,\la{Psi0}
\ee 
where the dots refers to terms that are either of cubic or higher order  in $C,\bar{C}$ or to terms purely quadratic in the same helicity fields  $\bar{C}$. In that sense the higher spin charges $\cQ_s$ that we study in the following are truncations of the Weyl tensor expansion coefficients for spin higher than $4$.\footnote{The results of \cite{Barnich:2019vzx} support our statement for $s=4$.}
 The fact  that the Weyl tensor coefficients are not fully determined from the higher spin charges, for $s\geq 4$, is likely  related to a puzzle in celestial holography appearing at spin-$4$ (see Section \ref{ss:celestial}).  
We leave the  precise relation between \eqref{dotQss} and the vacuum Einstein's equations for spin $s\geq 4$ to further studies.

\subsection{Linearized Einstein equations}\la{sec:LEE}

We now give a direct proof  that the $\Psi_0$ expansion \eqref{Psi0}   together with the evolution equations
\eqref{dotQss} truncated to linear order in the shear field (i.~e. keeping only the spatial derivative term on the RHS)   allow us to precisely recover 
the full content of the  linearized Einstein vacuum theory at all orders in the large-$r$ expansion around null infinity.
In order to do this, we rely on the analysis of Newman and Penrose  in \cite{Newman:1968uj} (in Section \ref{sec:w-structure} we comment on the relation with the set of conserved charges  introduced there).

Given the  asymptotic expansion of the $\Psi_0$ Weyl scalar 
\be
\Psi_0 &=\sum_{n=0}^\infty \f{\Psi_0^{(n)}}{r^{5+n}}\,,
\ee
at linear order the Bianchi identities imply the following set of evolution equations \cite{Newman:1968uj}
\be
\dot \Psi_0^{(n+1)}&= -\f1{  (n+1)} \left(\bar D D +\frac12 n(n+5) \right) \Psi_0^{(n)}
\,.\la{dP0n}
\ee
To compare with existing literature we evaluate the operators $D$ and $\bar{D}$  in complex coordinates on the round sphere.
For the round sphere metric $\rd s^2 =\frac{2 \rd z\rd \bar{z}}{P^2}$, and we have  the complex  frame   $\hat{m} :=m^A\pa_A= {P} \pa_{z}$,
where $P:= \frac{(1+z\bar{z})}{\sqrt{2}} $. As shown in Appendix  \ref{app:edth} we find that 
\be
D O_s = P^{1-s} \pa_z (P^s O_s), \qquad \bar{D} O_s = P^{1+s} \pa_{\bz} (P^{-s} O_s),
\ee
which shows that these are proportional to the {\it edth} differential operator on the sphere \cite{Goldberg, eastwood_tod_1982,  doi:10.1063/1.528587}.  In particular we have $\eth= \sqrt{2} D$ and $\bar\eth=\sqrt{2} \bar{D}$.
These expressions imply that 
\be
[ \bar{D}, D] O_s = s O_s.\label{comm}
\ee
It is important to remember that $D$ raises the spin by one unit while $\bar{D}$ lowers the spin by one unit.
The sphere Laplacian acting on spin $s$ observables can be diagonalized in terms of spin $s$ spherical harmonics $Y^s_{\ell,m}$ of angular momenta $\ell\geq |s|$. The eigenvalues of the Laplacian  are given by 
\be\la{ethdiag}
 \bar{D} D Y^s_{\ell,m}= -\frac12 (\ell-s)(\ell+s+1) Y^s_{\ell,m}.
\ee
This means that the set of fields of helicity $s$ can be decomposed as 
\be 
V^s = \oplus_{\ell=s}^\infty V^s_\ell,
\ee
where $V^s_\ell$ is of dimension $2\ell+1$ and it is spanned by the higher spin spherical harmonics $Y^s_{\ell,m}$.
We can use this to decompose 
\be\la{Psi0}
\Psi_0^{(n)}= \Psi_{G0}^{(n)} +\Psi_{L0}^{(n)}\,,
\ee
 where 
$\Psi_{G0}^{(n)} $ is the \emph{global} component of $\Psi_0$, while 
$\Psi_{L0}^{(n)}$ is its \emph{local} component.
Both components are spin $2$ fields, what differentiates them is the fact that the global component only contains angular
momenta of  value  $\ell=\{2,\cdots, n+1\}$ while the local component can be decomposed in terms of fields with angular momenta 
$\ell\geq 2+n $. Explicitly,
this implies the expansion
\be
 \Psi_{G0}^{(n+1)} = \sum_{k=0}^{n} 
 \left[\Psi_{0}^{(n+1)}\right]_{\ell=2+n-k}\,, \quad \mathrm{with} \quad  \left[\Psi_{0}^{(n+1)}\right]_{\ell=2+n-k}  \in V^2_{2+n-k}\,,
\la{Psi0G}
 \ee
 where $\left[\Psi_{0}^{(n+1)}\right]_{\ell}$  is the projection of $\Psi_0$ onto $V^{2}_\ell$.
Similarly
\be
\Psi_{L0}^{(n+1)} = \sum_{k=0}^{\infty} 
 \left[\Psi_{0}^{(n+1)}\right]_{\ell=2+n+k}\,, \quad \mathrm{with}\quad  \left[\Psi_{0}^{(n+1)}\right]_{\ell=2+n+k}  \in V^2_{2+n+k}.
\ee
Both components satisfy \eqref{dP0n} since there is no mixing at the linear level.

Now since the operator $ \bar{D}^n$ maps $V^{2+n}=\oplus_{\ell=2+n}^\infty V^{2+n}_\ell$ onto $V^{2}_L:=\oplus_{\ell=2+n}^\infty V^2_\ell$, the local component is in the image of the map $ \bar{D}^n$ acting on spin $s = n + 2$ fields. This means that we can express this component 
in terms of the the higher spin charges, namely 
\be\la{Psi0L}
\Psi_{L0}^{(n)} &=  \frac{(-)^n}{n ! }\bar{D}^{n} \cQ_{n+2} \,,  \qquad {\rm for} \qquad n>0\,.
\ee
We thus see that the evolution equation \eqref{dP0n}  becomes
 \be
\bar{D}^{n+1} \dot \cQ_{n+3} &= \left(\bar D D  + \frac{1}{2} n(n+5) \right)\bar{D}^{n} \cQ_{n+2} = \bar{D}^{n+1} D \cQ_{n+2}\,,
 \ee
 where we used the commutator \eqref{comm} to evaluate
 \be
\bar{D} D \bar{D}^{n} \cQ_{n+2}&= -\left( \sum_{\ell=3}^{n+2} \ell \right) \bar{D}^{n} \cQ_{n+2}+   \bar{D}^{n+1} D \cQ_{n+2}\cr
&=-\frac12 n(n+5) \bar{D}^{n} \cQ_{n+2}+   \bar{D}^{n+1} D  \cQ_{n+2}\,.
\ee 
The equation ${\cal E}_s := \dot  \cQ_{s+1}- D \cQ_s$ takes place in $V^{s+1}$ and the map $\bar{D}^{s-1}: V^{s+1} \to V^{2} $  is injective. 
Therefore,  the linearized time-development Bianchi identities yield the charge evolution equations 
\be
\dot  \cQ_{s+1}= D \cQ_s \,,  \qquad {\rm for} \qquad s\geq 2\,,
\ee
corresponding to the linear version of our recursion relation
\eqref{dotQss}.
The equations for the spins $s=-2,-1,0,1$ can be obtained from the linearization of the Bianchi identity applied respectively to
$\Psi_4^{(0)},\Psi_3^{(0)}, \Psi^{(0)}_2$ and $\Psi_1^{(0)}$.
 This means that \eqref{dotQss} captures the full content of Einstein's equations in the linearized theory for all spins, namely at all orders in the large-$r$ expansion around null infinity.

We are left with the analysis of the global component $\Psi_{G0}^{(n+1)}$ which contains the Newman-Penrose global charges.
The Newman-Penrose  charges \cite{Newman:1968uj} are  given as the $\ell = 2+n$ component of $\Psi_{G0}^{(n+1)}$.
 They are denoted by $G_n:=\left[\Psi_{0}^{(n+1)}\right]_{\ell= 2+n} \in V^2_{2+n}$.
From the evolution equation \eqref{dP0n} and the fact that $\left.\bar{D}D\right|_{V^2_{2+n}}=-\frac12 n(n+5)$ one gets that they are conserved in time.
More generally the global components are polynomial in the time $u$ and satisfy 
\be
\pa_u^n  \Psi_{G0}^{(n)} = 0,
\ee
which follows directly from the evolution equation, and the eigenvalue equation 
\be 
\prod_{k=1}^{n} \left[ \bar{D} D +\frac12 k (k + 5)\right]  
Y^2_{2+p,m}=0, \quad \mathrm{for}\quad p\leq n.
\ee
As shown in the Appendix \ref{app:edth}, these constants uniquely determine, through the evolution equation, the polynomials $\Psi_{G0}^{(n+1)}$:
\be
\label{eq:gl-sol}
 \Psi_{G0}^{(n+1)}(u) =\sum_{k=0}^n \alpha_{n}^k   G_{k}\, u^{n-k} ,
\ee
where 
$\alpha_n^k$ are given by 
\be
\label{eq:gl-alpha}
\alpha_n^{k} =  \frac{  (-1)^{n-k}}{2^{n-k} } \frac{ (k+1)!}{(n+1)!} \frac{ (n+k+5)!}{(2k+5)!}.  
\ee

Finally, an important point to appreciate is that $\Psi_0$ captures, in the Bondi gauge, information about the radial expansion of the sphere metric. In particular, an expansion $\gamma_{AB}(r)= q_{AB} -\frac1r C_{AB} +\sum_n r^{-n} q_{AB}^{(n)}$ implies 
$ \Psi_0^{(n)}\propto  q^{(n+3)}+\cdots $.
Moreover, the radial Einstein equation $G_{\langle AB\rangle }=0$, with $G_{\mu\nu}$ the Einstein's tensor,  implies that the $r$ dependence of $\pa_u \Psi_0$ is determined by its value at $r=\infty$. On the other hand, the values of $\Psi_0$ at any cut $u=$ cst. are free data from the point of view of $\scri^+$. We expect these free data to be encoded into the higher spin charges $\cQ_s$ for $s\geq 2$.

\subsection{Higher spin symmetry action}
\label{ss:hssa}

Having provided some motivation for considering the recursion relations \eqref{dotQss}, we now study their implications for the symmetry algebra of null infinity.
Substituting \eqref{charge-kexp} into \eqref{highQ} and equating terms with the same number of oscillators, we find a recursion relation at each order $k$, 
\be
\la{charge-k-rec}
\cQ_{s}^{k} &= D\pa_u^{-1} (\cQ_{s-1}^{k}) + \frac{(s+1)}{2} \pa_u^{-1}(C \cQ_{s-2}^{k-1}).
\ee
Recalling that
\be 
\mathcal{Q}_{-2} = \mathcal{Q}_{-2}^{1} = \frac{1}{2} \mathcal{N}\,,
\ee
\eqref{charge-k-rec} can be solved order by order in $k$ for any $s \geq -1$. We present the explicit solution for the first two orders $k=1,2$. 
For the soft charge ($k = 1$), the second term in \eqref{charge-k-rec} drops out and we simply find
\be 
\la{Q1}
\cQ_{s}^{1}(u,z) =(\p_u^{-1} D)^{s+2} \cQ_{-2}^1(u,z) = \frac12  (\p_u^{-1} D)^{s+2}\cN(u,z).
\ee 
This result can be used to evaluate the quadratic ($k = 2$) contribution for $s\geq 0$
\be\la{Q2}
\cQ_s^{2}(u,z)&= \frac14
\sum_{\ell=0}^s (\ell+1) \p_u^{-1}(\p_u^{-1} D)^{s-\ell} \left[C  (\p_u^{-1} D)^{\ell} \cN  \right](u,z).
\ee 
As explicitly shown in \cite{Freidel:2021dfs} for the cases $s=1,2$, the action of the charges $\cQ_{s}$ on $C$ leads to divergent contributions when $u\to-\infty$ and a renormalization procedure is required. Remarkably, as noted in \cite{Freidel:2021dfs} this renormalization yields charges parametrizing  the non-radiative corner phase space \cite{Freidel:2021qpz}, meaning the charges are conserved in time when the no radiation conditions $\cN=0=\cJ$ are imposed. Generalizing the renormalization procedure of \cite{Freidel:2021dfs}
 to all $s$, we define the renormalized higher spin generators
 \be\la{Qren}
 \hat q_s(u,z):=  \sum_{n=0}^{s} \frac{(-u)^{s-n}}{(s-n)!} D^{s-n} \cQ_n(u,z)\,.
\ee
The higher spin charge aspects are then obtained as the limit\footnote{Note that here and in the following we use the short-cut notation $F(z)$ to denote a function on the sphere implicitly taken to depend on both coordinates $z, \bz$ on the sphere. We do not imply that $F$ is holomorphic. When explicitly needed, we restore the dependence on both coordinates.}
\be\la{Qren2}
q_s(z)=\lim_{u\to-\infty} \hat q_s(u,z).
\ee 
This limit is now well defined under the assumption  \eqref{Nbc}.
 We next  separately analyze the action of the renormalized linear and quadratic higher spin generators on the gravitational phase space variables.

\subsubsection{Higher linear generators}\la{sec:Hlinear}

 The identity\footnote{This follows from the generalized Leibniz rule \be 
 \pa_u^{-1} (fg)=\sum_{n=0}^{\infty} (-1)^n (\pa_u^n f)\pa_u^{-(n+1)}g.
 \ee }
\be 
\pa_{u}^{-1}\left( \frac{u^k}{k!} f(u)\right)
&=(-1)^k\sum_{n=0}^{k}\frac{(-u)^{(k-n)}}{(k-n)!}
\pa_{u}^{-(n+1)}f(u)\la{Leibn}
\ee
allows us to relate the the $k=1$ contribution to the renormalized higher spin corner charge aspects \eqref{Qren2} to a negative-helicity soft graviton mode, namely
\be\la{Q1ren}
 q^1_s(z)&:= \lim_{u\to-\infty} \sum_{n=0}^{s} \frac{(-u)^{s-n}}{(s-n)!} D^{s-n} \cQ^1_n(u,z)
= D^{s+2}N_s(z).
\ee
Here we have introduced the (negative helicity) (sub)$^s$-leading soft graviton operator 
\be\la{Nsdef}
N_s(z) :=\f{1}2\f{(-1)^{s+1}}{s!}\int_{-\infty}^\infty \rd u\, u^s \hat N(u,z)\,.
\ee
$N_s$ can be expressed in terms of modes upon defining the Fourier transform
\be 
\label{Nmode}
\begin{split}
N^{\omega}(z) := \int_{-\infty}^{\infty} \rd u\, e^{i\omega u} \hat N(u,z)\,.
\end{split}
\ee
Then
\be
N_s &= -\f14 \frac{(-i)^s}{s!}\lim_{\omega \rightarrow 0^+}({-}\p_{\omega})^s\left(N^{\omega}+ (-1)^s N^{-\omega} \right)\cr
&= \frac{\kappa}{16\pi}  \frac{(-i)^s}{s!} \lim_{\omega \rightarrow 0^+}(\p_{\omega})^{s-1}(1+\omega\p_{\omega})\left(a_{ +}^{\rm out \dagger}(\omega \hat{x}) + (-1)^s a_{ -}^{\rm out}(\omega  \hat{x}) \right)\,,\la{Ns}
\ee
where in the last line we used the mode expansion \eqref{Nexp}. One can check that for $s=0,1,2,$ \eqref{Ns} reduce to the known expressions for the leading, subleading, and sub-subleading soft charges \cite{Kapec:2016jld, Campiglia:2016efb}.

The definition \eqref{Q1ren} extends the  result of \cite{Freidel:2021dfs} to all higher spins and relates the higher spin soft charges to soft graviton modes. Note that since $D$ is an operator of dimension/helicity $(1,1)$, the relation $q^1_s(z)=  D^{s+2}N_s (z)$ implies that $N_s$ has dimension/helicity $(\Delta, J)= (1-s,-2)$.\footnote{We can also establish this directly since  $\hat N(z)$ has dimension/helicity  $(\Delta, J)= (2,-2)$, and $\rd u u^s$ has $(\Delta, J)= (-s-1,0)$.}

\subsubsection{Higher quadratic generators}\la{sec:Hquadratic}

For the quadratic contribution $k=2$, the renormalized expression takes the form
\be\la{Q2ren}
\hat q^2_s(u,z)
&= \frac14\sum_{n=0}^{s}\sum_{\ell=0}^n \frac{(\ell+1) (-u)^{s-n}}{(s-n)!} 
\p_u^{-(n-\ell+1)}  D^{s-\ell}\left[C  (\p_u^{-1} D)^{\ell} \cN  \right](u,z)\,.
\ee
Together with \eqref{NCbra}, this allows us to compute the action of $\hat q_s^2$ on $C$, \footnote{Here and in the following, the subscripts $z,z'$ in the covariant derivative are added just to keep track of the quantities they act upon, when this is necessary. They do not represent  spatial indices.}
\be
\la{quadr-action}
\{\hat q_s^{2}(u,z), C(u', z')\} 
   &= \f{\kappa^2}8 \sum_{n=0}^{s}\sum_{\ell=0}^n  \frac{(-u)^{s-n}}{(s-n)!}  
   (\ell+1)\cr
   &\times \p_u^{-(n-\ell+1)}\left[D^{s-\ell}_z\left(C(u,z)D_z^{\ell} \delta(z,z')  \right)   \left(\p_u^{-(\ell-1)} \delta(u-u')\right)\right]\cr
   &= \f{\kappa^2}8 \sum_{n=0}^{s}\sum_{\ell=0}^n  \frac{(-)^{\ell}(-u)^{s-n}}{(s-n)!}  
   (\ell+1) \cr
   &\times \p_{u'}^{-(\ell-1)}\left[D^{s-\ell}_z\left(C(u',z)D_z^{\ell} \delta(z,z')  \right) \f{ (u-u')^{n-\ell}}{(n-\ell)!}  \theta(u'-u)\right]\,,
\ee
where  we have used
the identities
\be \la{id}
\begin{split}
\pa_u^{-a} f(u) \pa_u^{-b} \delta(u - u') &= (-1)^b \p_{u'}^{-b}f(u') \pa_u^{-a} \delta(u - u'), \\
\p_{u}^{-k} [f(u) \delta(u-u')]& = -\f{(u-u')^{k-1}}{(k-1)!} f(u')\theta(u'-u)\,.
\end{split}
\ee
The second one follows by recurrence  from our definition \eqref{iiint}.

Switching the order of the sums and using that 
\be 
\sum_{n=\ell}^{s} \f{ (-u)^{s-n} (u-u')^{n-\ell}}{(s-n)!(n-\ell)!} =\frac{ (-u')^{s-\ell}} {(s-\ell)!},
\ee
\eqref{quadr-action} becomes
\be
\{\hat q_s^{2}(u,z),& C(u', z')\} 
    = \f{\kappa^2}8 \sum_{\ell=0}^s  (-)^{\ell}
   (\ell+1) 
\p_{u'}^{-(\ell-1)}\left[D^{s-\ell}_z\left(C(u',z)D_z^{\ell} \delta(z,z')  \right)\theta(u'-u) \f{ (-u')^{s-\ell}} {(s-\ell)!}  \right]\cr
&= \f{\kappa^2}8 \sum_{\ell=0}^s  \sum_{n=0}^\ell  (-)^{s+n}
   \frac{(\ell+1)!}{n!(\ell-n)!} 
\p_{u'}^{-(\ell-1)}\left[\left(D_{z'}^{n}C(u',z')D_z^{s-n} \delta(z,z')  \right) \f{ u'^{s-\ell}\theta(u'-u)} {(s-\ell)!}  \right]\,.
\ee
The last equality follows from
\be 
\begin{split}
C(z) D_z^{\ell}\delta(z, z') 
&= \sum_{n = 0}^{\ell} (-1)^{n} \left(\begin{matrix}
\ell\\
n
\end{matrix}\right) D_{z'}^n C(z') D_{z}^{\ell - n} \delta(z, z').
\end{split}
\ee
Note that the action of the renormalized charge is manifestly finite in the limit $u \rightarrow -\infty$, which we can now take.

A final simplification occurs using the Leibniz rule from pseudo-differential calculus  \cite{Gelfand-Dickey, Adler, DickeyB, Bakas:1989um, Bonora:1994fq} generalizing \eqref{Leibn}, 
\be \la{Leibn2}
\pa_{u'}^{\alpha}\left( \frac{u'^k}{k!} C(u')\right)= \sum_{n=0}^{k} \frac{(\alpha)_n}{n!}\frac{u'^{(k-n)}}{(k-n)!}
\pa_{u'}^{\alpha -n }C(u') = \frac{1}{k!}(\Delta + \alpha - 1)_{k} \p_{u'}^{\alpha - k}C(u'),
\ee 
 where the last equality can be proven by recurrence on $k$ and we defined 
\be 
\label{delta-u}
\Delta - 1 := u'\p_{u'}\,,
\ee
while $(x)_n = x(x - 1)...(x - n + 1)$ is the falling factorial. Then\footnote{See Appendix \ref{App:Norm} for a more direct proof in a conformal primary basis.}
\be \la{pip}
\p_{u'}^{-(\ell-1)} \left(C(u') \frac{u'^{s - \ell}}{(s - \ell)!} \right) = \frac{(\Delta - \ell)_{s - \ell}}{(s - \ell)!} \p_{u'}^{1-s} C(u')\,, 
\ee
and we conclude 
\be
\{q_s^{2}(z), C(u', z')\} &= 
\f{\kappa^2}8\sum_{\ell=0}^s\sum_{n=0}^\ell (-)^{s+n}   \f{(\ell+1)!}{n!(\ell-n)!}  \f{\left (\Delta-\ell \right)_{s-\ell}}{(s-\ell)!}  \p_{u'}^{1-s} D^n_{z'} C(u',z')  D_z^{s-n} \delta(z,z')\,.
\la{r-chs-a}
\ee
Evaluating the sum over $\ell$ first (see proof in Appendix \ref{app:CB}),
\be \la{DD}
\sum_{\ell = n}^{s} \frac{(\ell + 1)! (\Delta - \ell)_{s-\ell}}{(\ell - n)! (s-\ell)!} 
=  \frac{(n+1)!}{(s-n)!} (\Delta + 2)_{s - n},
\ee
the bracket  \eqref{r-chs-a} becomes
\be 
\label{bulk-s-cfin}
\boxed{\,
\{q_s^{2}(z), C(u', z')\} = \frac{\kappa^2}{8}\sum_{n = 0}^s (-1)^{s + n}
\frac{(n+1)(\Delta + 2)_{s - n}}{(s-n)!} \p_{u'}^{1-s} D_{z'}^n C(u', z') D_{z}^{s - n} \delta(z, z'). \,}
\ee
For the opposite helicity, we find after a similar analysis, presented in Appendix \ref{App:barC}, that 
\be
\label{bulk-s-cbfin}
\boxed{\,
\{q_s^{2}(z), \bar C(u', z')\}
=\f\k8\sum_{n=0}^s (-1)^{s+n} \f{(n+1)(\Delta -2)_{s-n} }{(s-n)!}  \p_{u'} ^{1-s}D^n_{z'} \bar C (u',z') D_z^{s-n}\delta(z,z') \,.
}
\ee
These equations determine the action of the quadratic spin-$s$ charge on the gravitational phase space.
They generalize the actions of the complex mass $m_\C=\frac{8}{\kappa^2} q_0$, momentum $p=\frac12\frac{8}{\kappa^2} q_1$ and spin-$2$ charge $t=\frac13\frac{8}{\kappa^2} q_2$ worked out in \cite{Freidel:2021dfs}.

The actions \eqref{bulk-s-cfin} and \eqref{bulk-s-cbfin} allow us to straightforwardly evaluate the brackets of the quadratic charges with soft gravitons. In particular, for negative-helicity soft gravitons, using the definition \eqref{Nsdef} and the bracket \eqref{bulk-s-cbfin}, we find that  (see Appendix \ref{App:SGbra} for a detailed derivation)
\be 
\la{quadr-soft-gr-action}
\{q^2_s(z), N_{s'}(z')\} &=\f{(-1)^{s'+1}}2\f1{s'!}\int_{-\infty}^\infty \rd u\, u^{s'} \{q^2_s(z),\hat { N}(u,z')\}
\cr
&= \frac{\kappa^2}{8}\f{(-1)^{s'+s+1}}2 \sum_{n = 0}^s  
\frac{(-1)^{ n} ( n+1)}{s'!(s-n)!} D_{z}^{s - n} \delta(z, z') \cr
&\times  D_{z'}^n \int_{-\infty}^\infty \rd u\,  u^{s'} \p_u(\Delta - 2)_{s - n} \p_{u}^{1-s}\bar{C} (u, z') \,.
\ee 
The terms inside the integral can be rearranged to give\footnote{ The next equality uses the definition ${\hat N}:=\pa_u \bar{C}$ given above, which is valid only in the spherical metric frame where $\hat{N}=N$.
}
\be
\f12 \int_{-\infty}^\infty \rd u\,  u^{s'} \p_u(\Delta - 2)_{s - n} \p_{u}^{1-s} \bar{C} (u, z') 
&=\f12 \int_{-\infty}^\infty \rd u\,  \frac{(\Delta -s' -1)_{s -n}}{(\Delta-s'-1)_{s-1}} u^{s+s'-1} \hat N (u, z')\cr
&= (-1)^{s+s'+ n +1}(s+s'-n)! 
N_{s+s'-1}.
\ee 
In the last equality we have used that the operator $\Delta=\pa_u u$ integrates to $0$. Substituting this into \eqref{quadr-soft-gr-action}, we conclude that
\be\la{q2Ns}
\{q^2_s(z),  N_{s'}(z')\}&=
\frac{\kappa^2}{8} \sum_{n = 0}^s(n+1) 
\left(\begin{matrix}
s+s'-n\\
s'
\end{matrix}\right)
(D_{z'}^n  N_{s+s'-1}(z')) D_{z}^{s - n} \delta(z, z')\,.
\ee
In Section \ref{sec:tower} we show that \eqref{bulk-s-cfin} and \eqref{bulk-s-cbfin} reproduce the action of the infinite tower of conformally soft symmetries implied by the celestial OPE block  \cite{Guevara:2021abz}, while \eqref{q2Ns} is equivalent to the special case when both gravitons in the OPE are taken to be soft.
We conclude our analysis of the (truncated) charge action on phase space by showing that this can be written entirely in terms of the action of a pseudo-differential operator, generalizing to higher spins one of the central results of \cite{Freidel:2021dfs}.

\subsection{Higher spin pseudo-differential operators}\la{sec:pseudo}

In this section we show that the spin-$s$ quadratic charge action is implemented on the gravity phase space by the action of a pseudo-differential operator.
We recall that according to \cite{Freidel:2021dfs} a pseudo-vector  of spin $p$ is an operator  of dimension/spin $(1,p)$ given by ${\cal{D}}_p:=D_z^p\pa_u^{1-p}$.

Integration of the higher spin charge aspects against a function $\tau_s(z)$ on the sphere yields the higher spin charges \footnote{ $q$ is the determinant of the leading order 2-sphere metric $\gamma_{AB}$ in asymptotically flat metrics \eqref{ds}.}
\be\la{Qs}
Q_s(\tau):=\f8\k\int_{S}\rd^2z \sqrt{q}\, \tau_s(z) q_s(z)\,.
\ee
The action of the quadratic component of these charges on $C$ is then given by  
\be 
\{Q_s^{2}(\tau), C(u, z')\}
= \sum_{p = 0}^s \frac{u^{s-p}}{(s-p)!} \delta^p_{D^{s - p}\tau_s} C(u, z')\,,
\ee
where $\delta^p_{\tau_p}$
is the action of a spin-$p$ pseudo-vector field  on $C$. This takes the form
\be \la{dtauC}
\delta_{\tau_p}^p C :=
\sum_{k=0}^{\mathrm{min}[3, p]} 
 \left(\begin{matrix}
3\\
k
\end{matrix}\right)(p+1-k)
 (D^{k}\tau_p) \,  [D^{p-k}\pa_u^{1-p} C].
\ee 
Note that this action is such that $\delta_{\tau_p}^p C = (p+1) \tau_p {\cal D}_p C + \cdots$ where the dots denote tensorial corrections. 
For low spin \eqref{dtauC} reduce to
\be 
\delta_{\tau_0}^0 C &=
\tau_0 \, \pa_u C,\cr
\delta_{\tau_1}^1 C &=
 2 \tau_1 \, D C + 3(D\tau_1) \,
  C,\cr
  \delta_{\tau_2}^2 C &=
 3 \tau_2 \,  D^{2}\pa_u^{-1} C
 + 6 (D\tau_2) \,  D\pa_u^{-1} C
 + 3D^2(\tau_2 \,  \pa_u^{-1} C).
\ee 
We recognize the action of the supertranslation, diffeomorphism and spin-$2$ transformations on the shear \cite{Freidel:2021dfs}. 

Similarly, the action of the charge on $\bar{C}$ is given by  
\be 
\{Q_s^{2}(\tau), \bar{C}(u, z')\}
= \sum_{p = 0}^s \frac{u^{s-p}}{(s-p)!} \delta^p_{D^{s - p}\tau_s} \bar{C}(u, z')\,,
\ee
where spin-$p$ pseudo-vector fields $\delta^p_{\tau_p}$ act on $\bar{C}$ as
\be \la{dtaubC}
\delta_{\tau_p}^p \bar{C} :=
\sum_{k=0}^{p} 
 \left(-1\right)^k (p+1-k)
 (D^{k}\tau_p) \,  [D^{p-k}\pa_u^{1-p} \bar{C}].
\ee 

These identities can be proven by starting with the expression 
\be 
\{Q_s^{2}(\tau), C(u, z')\} &= \sum_{n = 0}^s 
\frac{(n+1)(\Delta + 2)_{s - n}}{(s-n)!}  (D_{z}^{s - n}\tau_s)    D_{z'}^n \p_{u}^{1-s} C(u, z')
\ee 
and using the identity (see Appendix \ref{App:pseudo})
\be
\frac{(\Delta + 2)_{s - n}}{(s-n)!} = \sum_{k=0}^{\mathrm{min}[3, s-n]}\left(\begin{matrix}
3\\
k
\end{matrix}\right) \frac{u^{s-n-k}\pa_u^{s-n -k}}{(s-n-k)!}\,.
\ee 
Therefore, we find
\be 
\{Q_s^{2}(\tau), C(u, z')\}& = \sum_{n = 0}^s 
\sum_{k=0}^{\mathrm{min}[3, s-n]} \left(\begin{matrix}
3\\
k
\end{matrix}\right)
\frac{(n+1)u^{s-n-k}}{(s-n-k)!}  (D_{z}^{s - n}\tau_s)    D_z^n \pa_u^{1-n-k} C(u, z') \cr
&=\sum_{p = 0}^s \frac{u^{s-p}}{(s-p)!} \sum_{k=0}^{\mathrm{min}[3, p]} 
  \left(\begin{matrix}
3\\
k
\end{matrix}\right) (p+1-k)
 (D_{z}^{s - p +k}\tau_s)    [ D^{p-k}\pa_u^{1-p} C(u, z')] \cr
 &=\sum_{p = 0}^s \frac{u^{s-p}}{(s-p)!} \delta^p_{D^{s - p}\tau_s} C(u, z')   
\ee 
as anticipated.
The proof for the action on $\bar{C}$ is analogous and is given in Appendix \ref{App:pseudo}.

\section{Tower of soft theorems and celestial symmetries}\la{sec:tower}

In this section we connect the asymptotic symmetry analysis in the previous sections to the recently uncovered conformally soft theorems \cite{Pate:2019mfs, Pate:2019lpp, Puhm:2019zbl, Guevara:2021abz}.
In particular, we derive in Section \ref{sec:soft-th}  the Ward identities arising from the conservation of all higher spin charges truncated to quadratic order in  $\kappa$, that is neglecting all higher order collinear terms.
We then demonstrate that these conservation laws are equivalent to the tower of tree-level conformally soft graviton theorems revealed by celestial holography. 
Furthermore, we demonstrate in Section \ref{ss:celestial} that the quadratic action \eqref{bulk-s-cfin} remarkably reproduces the action of the infinity of celestial soft symmetries whose algebra was computed holographically in \cite{Guevara:2021abz}.
After clarifying the relationship between the soft graviton and the $w$-current in Sections \ref{sec:w-structure} and \ref{sec:wcurr},
we  argue in Section \ref{sec:recursion}  that the quadratic parts of $q_s$ provide a spacetime realization of the $w_{1+\infty}$ algebra identified in \cite{Guevara:2021abz, Strominger:2021lvk}.   Finally, we explicitly compute the higher spin charge bracket to linear order in Section \ref{sec:CB} and show that this yields a canonical representation of the $w_{1+\infty}$ algebra.

\subsection{From conservation laws to soft theorems}\la{sec:soft-th}

We can extend the analysis of the  leading, subleading and sub-subleading Ward identities \cite{Strominger:2013jfa, Kapec:2014opa, Campiglia:2016efb, Conde:2016rom,Freidel:2021dfs} to all higher spin charges $q_s$ truncated to quadratic order. The truncated  Ward identity takes the form\footnote{Antipodal matching is implicit.}
\be 
\la{spin-s-cons}
\langle {\rm out}|[q_s^1,\mathcal{S}]|{\rm in}\rangle = -\langle{ \rm out}| [q_s^2, \mathcal{S}]|{\rm in}\rangle.
\ee
Using \eqref{Q1ren}, \eqref{Ns},  as well as crossing symmetry 
\be 
\lim_{\omega \rightarrow 0^+} \p_{\omega}^s \left(\omega\langle{\rm out}| a_-^{{\rm out}}(\omega \hat{x}) S|{\rm in}\rangle\right) = (-1)^{s+1}\lim_{\omega \rightarrow 0^+} \p_{\omega}^s \left(\omega\langle {\rm out}|S a^{{\rm in}\dagger}_+(-\omega\hat{x})|{\rm out}\rangle\right)\,,
\ee
we have
\be\la{ST1}
\langle {\rm out}|[q_s^1,\mathcal{S}]|{\rm in}\rangle =
\frac{\kappa}{8\pi}  \frac{i^s}{s!} \lim_{\omega \rightarrow 0^+}(\p_{\omega})^{s}D^{s+2} \omega \langle {\rm out}| a_{ -}^{\rm out}(\omega  \hat{x}))\mathcal{S} |{\rm in}\rangle \,.
\ee
At the same time, replacing the bracket \eqref{bulk-s-cfin} with the  quantum commutator and using the mode expansion \eqref{mode-exp},
\be
[q_s^2(z), a_{\pm}^{\rm out}(\omega\hat{x}')]&=-i\frac{\kappa^2}{8}\sum_{\ell = 0}^s (-1)^{s + \ell}
\frac{(1 + \ell)(2h_\pm)_{s - \ell}}{\Gamma(1 - \ell + s)} (-i\omega)^{-s + 1} D_{z'}^\ell a_{\pm}^{\rm out}(\omega\hat{x}') D_{z}^{s - \ell} \delta(z, z')\,,
\ee
where
\be 
2h_{\pm} = -\omega\p_{\omega} \pm 2\,.
\ee
We refer the reader to Appendix \ref{App:barC} for the commutator with negative helicity modes.

The quadratic contribution to the charge conservation law thus yields
\be\la{ST2}
\langle {\rm out}|[q_s^2,\mathcal{S}]|{\rm in}\rangle =
i^s\frac{\kappa^2}{8}\sum_{k=1}^n \sum_{\ell = 0}^s (-1)^{s+\ell}
\frac{(1 + \ell)(2h_k)_{s - \ell}}{(s- \ell )!} (\epsilon_k \omega_k)^{-s + 1} D_{z}^{s - \ell} \delta(z, z_k)
D_{z_k}^\ell \langle {\rm out}|\mathcal{S} |{\rm in}\rangle
\,,
\ee
with $\epsilon_k=+1$ for outgoing particles and  $\epsilon_k=-1$ for incoming ones.

Hence, we  see that each conservation law \eqref{spin-s-cons} implies a  corresponding soft graviton theorem
\be
 D^{s+2} \left( \lim_{\omega \rightarrow 0}(\p_{\omega})^{s} \omega \langle {\rm out}| a_{ -}^{\rm out}(\omega  \hat{x})\mathcal{S} |{\rm in}\rangle\right)
  &+\kappa \pi\sum_{k=1}^n \sum_{\ell = 0}^s (-1)^{s+\ell}
(1 + \ell)(s)_\ell(2h_k)_{s - \ell} (\epsilon_k \omega_k)^{-s + 1} \cr
&\times D_{z}^{s - \ell} \delta(z, z_k) D_{z_k}^\ell \langle {\rm out}|\mathcal{S} |{\rm in}\rangle\stackrel{C}{=}0 \, ,\la{STs}
\ee
where the equality $\stackrel{C}{=}$ means modulo collinear terms.
The soft theorems associated with positive helicity soft graviton insertions can be obtained by considering the conjugate of the higher spin charges \eqref{Q1}, \eqref{Q2}.

One can check that for $s=0,1,2$ we recover the results of \cite{He:2014laa, Kapec:2014opa, Freidel:2021dfs}. In analogy to the sub-subleading soft theorem for the spin-2 charge, the full Ward identities for the higher spin charges contain collinear contributions and induce higher order {\it classical} corrections to the soft theorems \eqref{STs} up to ${\cal O}(\kappa^s)$ for a given spin-$s$ charge. The precise form of these collinear terms in the case $s=2$ has been derived in \cite{Freidel:2021dfs}.  We leave the computation of these corrections for $s>2$ to the future.

\subsection{Recovering the celestial soft symmetries}
\label{ss:celestial}

 It is natural to suspect that the conservation of the higher spin charges \eqref{dotQss} truncated to quadratic order is related to the infinite tower of (tree-level) soft symmetries of the $\mathcal{S}$-matrix found in \cite{Guevara:2021abz}. In this section we show that this is indeed correct.

As shown in the previous section, the left hand side of \eqref{spin-s-cons} corresponds to a soft insertion at $\mathcal{O}(\omega^s)$. 
Computing the right hand side by explicitly taking  the $\mathcal{O}(\omega^s)$ soft limit of the scattering amplitude with a graviton insertion is cumbersome using standard amplitudes techniques. Nevertheless, it was recently realized that celestial holography---a framework in which scattering observables are re-expressed in a basis of asymptotic boost rather than the conventional energy-momentum eigenstates---allows one to make a prediction about the tree-level behavior of arbitrarily subleading soft graviton insertions.
The main tool used in this argument is the celestial OPE \cite{Fan:2019emx, Pate:2019lpp, Guevara:2021tvr, Guevara:2021abz} 
of conformal primary gravitons $G^{\pm}_{\Delta}$. These can be represented as operators of dimension/helicity $(\Delta,\pm 2)$ and are simply given by 
\be 
\label{conf-gr}
G_{\Delta}^-(z) &:= -\frac{\Gamma(\Delta -1)}{2} \int_{-\infty}^{+\infty} \!\rd u\, u^{-\Delta +1} \hat{N}(u,z), 
\cr
G_{\Delta}^+(z) &:=  -\frac{\Gamma(\Delta -1)}{2} \int_{-\infty}^{+\infty} \!\rd u\, u^{-\Delta +1} \hat{\bar{N}}(u,z).
\ee 
Note that since $\hat{N}$ (resp. $\hat{\bar{N}}$) is 
of dimension/helicity $(2,2)$ (resp. $(2,-2)$) and $u$ is of dimension/helicity $(1,0)$, $G_\Delta^\pm$ indeed have the expected dimension/helicity $(\Delta, \pm 2)$.
It is convenient to express this operator in terms 
of $\hat{N}$ and $\hat{\bar N}$, since these satisfy the asymptotic conditions \eqref{Nbc}.
Of crucial importance will be the fact that the residues of $G_\Delta^{-}$ at negative integer dimensions are precisely the (sub)$^s$-leading soft graviton modes \eqref{Nsdef},
\be 
\label{cs-def}
\mathrm{Res}_{\Delta= 1-s}\left( G_\Delta^-(z) \right)= N_s(z) = \frac{(-1)^{s+1}}{2 {s!}}\int_{-\infty}^{+\infty} \!\rd u\, u^{s} \hat{N}(u,z).
\ee
In Appendix \ref{App:Norm} we show that the conformal gravitons defined in \eqref{conf-gr} are proportional to conformal primary boost eigenstates denoted by $O_{\Delta}^\pm$ \cite{Pasterski:2016qvg, Pasterski:2017kqt} and related to asymptotic on-shell graviton states by a Mellin transform 
\be 
|p(\omega, z, \bz)\rangle \rightarrow |\Delta, z, \bz \rangle = \int_0^{\infty} d\omega \omega^{\Delta -1} |p(\omega, z, \bz)\rangle.
\ee
The relationship is 
\be 
\label{propOG}
O_{\Delta}^\pm = i^{\Delta} \frac{8\pi}{i\kappa} G_{\Delta}^\pm.
\ee

For simplicity, in this and the following sections we work in coordinates where the celestial sphere is flattened to a plane (the conventions are summarized for example in \cite{Pate:2019lpp,Freidel:2021dfs}). 
One can compactly express the behavior of two gravitons in the antiholomorphic collinear limit.\footnote{Such a limit can be taken in bulk $(2,2)$ signature where $z, \bz$ are real independent variables.} One finds \cite{Pate:2019lpp, Guevara:2021tvr}
\be 
\label{OPEs}
\begin{split}
O_{\Delta_1}^-(z_1) O_{\Delta_2}^{\pm}(z_2) \sim -\frac{\kappa}{2} \frac{1}{\bz_{12}} \sum_{n = 0}^{\infty} B(\Delta_1 - 1 + n, 2h_{2\pm} + 1) \frac{z_{12}^{n + 1}}{n!} \p^n O^{\pm}_{\Delta_1 + \Delta_2}(z_2) + \mathcal{O}(\bz_{12}^0)\,, 
\end{split}
\ee
with $	  z_{12}=z_1-z_2, ~ \bz_{12}=\bz_1-\bz_2$, and
where $2h_{2\pm} = \Delta_2 \pm 2$ and $J_2 = \pm 2$  for positive and negative helicity gravitons respectively; we have also introduced the Euler beta function $B(x,y)$.
These expansions resum the contribution from a conformal primary and all its SL$(2, \mathbb{R})_L$ descendants \cite{Ferrara:1972uq} and can also be derived from symmetry arguments as shown in \cite{Himwich:2021dau}. In particular, the leading term (the primary) is determined by the soft-collinear behavior of scattering amplitudes, while  the infinity of (spinning) descendant contributions is required by Lorentz symmetry.
Similarly, in the holomorphic collinear limit one finds
\be
\label{+gOPE}
\begin{split}
O_{\Delta_1}^+(z_1) O_{\Delta_2}^{\pm}(z_2)\sim -\frac{\kappa}{2} \frac{1}{z_{12}} \sum_{n = 0}^{\infty} B(\Delta_1 - 1 + n, 2\bar{h}_{2\pm} + 1) \frac{\bz_{12}^{n + 1}}{n!} \bar{\p}^n O^{\pm}_{\Delta_1 + \Delta_2}(z_2) + \mathcal{O}(z_{12}^0)\,,
\end{split}
\ee
where $2\bar h_{2\pm} = \Delta_2 \mp 2$.

As anticipated in \eqref{cs-def}, negative-helicity conformally soft gravitons of dimension $\Delta = 1 - s$ are defined as \cite{Pate:2019mfs, Guevara:2021tvr}
\be 
\N_s(z_1) := \lim_{\Delta_1 \rightarrow 1-s}& (\Delta_1  + s - 1) G_{\Delta_1}^-(z_1), \quad s \geq 0,~ s\in \mathbb{Z}.
\ee
 In this limit, only a finite number of terms survive on the RHS of \eqref{OPEs} which amounts to the statement that conformally soft gravitons are organized into finite-dimensional $SL(2,\mathbb{R})_L$ representations of dimension $s+1$.
Defining the spin $s$ operators
\be 
\label{soft-symm}
q^{1}_{s}(z_1) := \lim_{\Delta_1 \rightarrow 1-s}(\Delta_1 + s - 1)\p_{z_1}^{2+s} G_{\Delta_1}^-(z_1) = \p_{z_1}^{2 + s} N_s(z_1)
\ee
and using \eqref{OPEs} we find that 
\be 
\label{gen-resum}
\begin{split}
 q^{1}_{s}(z_1) G_{\Delta_2}^{\pm}(z_2) \sim \frac{\kappa^2}{8i}  \sum_{n = 0}^{s} \frac{(-1)^{n - s}  (n + 1)}{(2h_{2\pm} + 1)B(1 + s - n, 2h_{2\pm} + 1 -s + n)} \p^{s - n}_{z_1} \delta^{(2)}(z_{12}) \p^n_{z_2} G_{\Delta_2 + 1 - s}^{\pm}(z_2)\,. 
\end{split}
\ee
 We have used
\be 
\lim_{\epsilon \rightarrow 0} \epsilon B(\epsilon + n - s, 2h_{2\pm} + 1) = \frac{(-1)^{n - s}}{( 2h_{2\pm} + 1)B(1 + s - n, 2h_{2\pm} - s + n + 1)}\,,
\ee
which follows from the  Euler's reflection formula for the gamma function
$
\Gamma(x) \Gamma(1 - x) = {\pi}/{\sin (\pi x)}.
$
It is straightforward to verify that in a conformal primary basis, \eqref{bulk-s-cfin} reduces to the RHS of \eqref{gen-resum} for $s= 0, 1, 2$ \cite{Freidel:2021dfs}.
To prove the equivalence between \eqref{gen-resum} and \eqref{bulk-s-cfin} for all $s$, we perform one final manipulation to  put \eqref{gen-resum}   into the form
\be 
\label{rec}
 q^{1}_{s}(z_1) G_{\Delta_2}^{{\pm}}(z_2) \sim 
\frac{ \kappa^2}{8i \,s!} \sum_{n = 0}^{s} (-1)^{ n -s}(2h_{2\pm})_{s-n} (s)_n (n+1)\p_{z_1}^{s - n}\delta^{(2)}(z_{12}) \p_{z_2}^n G_{\Delta_2+1-s}^{{\pm}}(z_2)\,.
\ee
If we consider a negative helicity soft graviton operator and set $\Delta_2=1-s'$, the OPE \eqref{rec}
implies\footnote{
This can easily be seen by  noticing
\be
(-1)^{ n -s}(2h_{2-})_{s-n} \f{(s)_n}{s!}
&=(-1)^{ n -s}\f{(-1-s')_{s-n} }{(s-n)!}
=\f{(s+s'-n)!}{(s-n)!s'!}\,.
\ee
}
\be
q^1_s(z) N_{s'}(z')&\sim
\frac{ \kappa^2}{8 i } \sum_{n=0}^s 
(n+1)\left(\begin{matrix}
s+s'-n\\
s'
\end{matrix}\right)
\p^n_{z'}N_{s+s'-1}(z') \p_z^{s-n} \d^{(2)}(z- z' ) \,.
\la{qNOPE}
\ee
Therefore, \eqref{rec} can be seen to be equivalent to the bracket \eqref{q2Ns}, explicitly
\be\la{OPE-bra}
q^1_s(z) N_{s'}(z') &\leftrightarrow   \frac{1}{i}
\{q^2_s(z),  N_{s'}(z')\}\,. 
\ee
The OPE for a positive soft graviton can be recovered from the analogous bracket of $q^2_s$ with $\bar{C}$ computed in Appendix \ref{App:barC}.
Similarly, upon defining 
\be 
\label{soft-symm-ah}
\bar{q}^{1}_{s}(z_1) := \lim_{\Delta_1 \rightarrow 1-s}(\Delta_1 + s - 1)\p_{\bz_1}^{2+s} G_{\Delta_1}^+(z_1)= \p_{\bz_1}^{2+s} \bar{N}_s(z_1),
\ee
one can  show that \eqref{+gOPE} implies 
\be 
\label{rec+}
\bar q^{1}_{s}(z_1) G_{\Delta_2}^{{\pm}}(z_2) \sim 
\frac{ \kappa^2}{8 i\, s!} \sum_{n = 0}^{s} (-1)^{ n -s}(2\bar{h}_2)_{s-n} (s)_n (n+1)\p_{\bz_1}^{s - n}\delta^{(2)}(z_{12}) \p_{\bz_2}^n G_{\Delta_2+1-s}^{{\pm}}(z_2).
\ee

To summarize, we have started from a pattern observed in a large-$r$ expansion of Einstein's equations and demonstrated it implies the infinity of soft symmetries identified independently, holographically in \cite{Guevara:2021abz}. Conversely, we could have started from the celestial OPEs \eqref{OPEs} implying the symmetry action \eqref{rec} and inferred the recursion relation \eqref{dotQss} for the higher spin charges. We find this perfect match, while perhaps expected, remarkable. It is prime evidence that celestial holography not only provides a new organizing principle according to symmetry, but also allows one to infer aspects of the asymptotic gravitational dynamics, which are otherwise (perturbatively) much harder to access. 

We conclude this section with a note of caution. From the celestial point of view, there is an important caveat which we have so far avoided by treating $z, \bz$ as independent real variables (corresponding to bulk theories analytically continued to $(2,2)$ signature). Our bulk analysis on the other hand pertains to standard Lorentzian backgrounds corresponding to Euclidean celestial theories in which $z$ and $\bz$ are complex conjugates. In this case, as explained in \cite{Pate:2019lpp}, the second OPE in \eqref{OPEs} also receives a contribution with a pole in $z_{12}$,
\be 
\label{corr-cope}
\begin{split}
    O_{\Delta_1}^-(z_1) O_{\Delta_2}^{+}(z_2) \sim & -\frac{\kappa}{2} \frac{1}{\bz_{12}} \sum_{n = 0}^{\infty} B(\Delta_1 - 1 + n, \Delta_2 +3) \frac{z_{12}^{n + 1}}{n!} \p^n O^+_{\Delta_1 + \Delta_2}(z_2) \\
    & -\frac{\kappa}{2}\frac{1}{z_{12}}\left(\bz_{12}B(\Delta_1+3,\Delta_2-1)O_{\Delta_1+\Delta_2}^-(z_2) + \mathcal{O}(\bz_{12}^2) \right) + \cdots,
\end{split}
\ee
where $\cdots$ denote terms regular in the limit $z_{12}, \bz_{12} \rightarrow 0.$
As $G^-_{\Delta_1}$  is taken conformally soft,  the $z_{12}^{-1}$ terms drop out as long as $\Delta_1 \geq -2$ since the OPE coefficients multiplying the $z_{12}^{-1}$ term are regular. This is precisely the order (up to and including $s = 3$) to which we could explicitly verify the recursion relation \eqref{dotQss}. As such, the celestial OPE \eqref{corr-cope} suggests that \eqref{dotQss} receives corrections beyond $s = 3$. Further corrections arise from higher dimension operators in the low-energy effective action \cite{Elvang:2016qvq,Jiang:2021ovh,Mago:2021wje}. We leave a complete understanding of this, as well issues arising when mixing helicity sectors \cite{Pate:2019lpp, Banerjee:2018fgd, Banerjee:2018gce, Banerjee:2020zlg, Banerjee:2021dlm, Guevara:2021tvr} to future work. 

\subsection{Celestial diamonds}
\label{sec:w-structure}

\begin{figure}[h!]
     \centering
        \begin{tikzpicture}[scale=1.3]
\begin{pgfonlayer}{foreground}
\draw[thick](0,2)node[above]{{\footnotesize $\tilde{q}_s :(-1, -s)$}} ;
\draw[thick,->](0,2)--node[above left]{{\footnotesize$\p_{\bz}^{2-s}$}} (-1+.05,1+.05);
\draw[thick,->](0,2)--node[above right]{{\footnotesize$\p_z^{2 + s}$}} (2-.05,0+.05);
\draw[thick,->] (-1+.05,1-.05)node[left]{ {\footnotesize$N_s:(1-s, -2)~~$} } --node[below left]{{\footnotesize$\p_z^{2 + s}$}} (1-.05,-1+.05) ;
\draw[thick,->] (2-.05,0-.05)node[right]{{\footnotesize $~~~ S[N_s]:(1+s,2)$}} --node[below right]{{\footnotesize$\p_{\bz}^{2-s}$}} (1.05,-1+.05) ;
\draw[thick](1,-1.2)node[below]{{\footnotesize$q_s(z,\bz):(3,s)$}};
\filldraw[black] (0,2) circle (2pt) ;
\filldraw[black] (2,0) circle (2pt) ;
\filldraw[black] (-1,1) circle (2pt) ;
\filldraw[black] (1,-1) circle (2pt) ;
\end{pgfonlayer}
\end{tikzpicture}
 \caption{Spin-$s$ diamond associated with negative-helicity soft gravitons and  $s =-1, 0, 1, 2$. Operators connected by long edges have weights $h=(\Delta+J)/2, \bar h=(\Delta-J)/2$ related by $(h, \bar{h}) \leftrightarrow (1 - h, \bar{h})$. Operators connected by short edges have $(h, \bar{h}) \leftrightarrow (h, 1 - \bar{h})$. Diagonally opposite corners are related by $(h, \bar{h}) \leftrightarrow (1 - h, 1 - \bar{h})$.}
    \label{fig:Diamond2}
\end{figure}
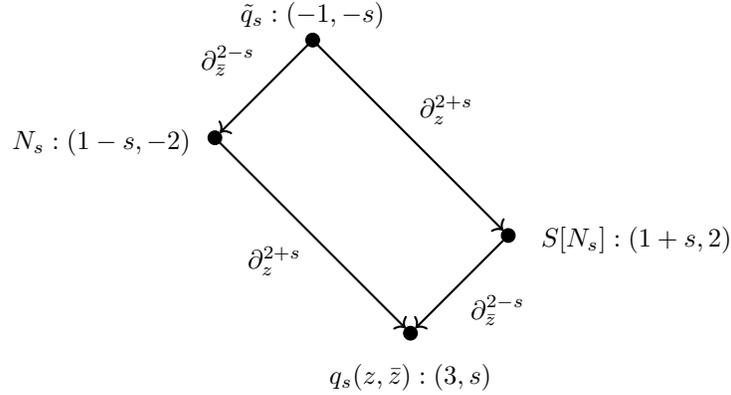

Conformal primary wavefunctions associated with the leading, subleading and sub-subleading soft gravitons (i.e. with $\Delta = 1, 0, -1$ and  $J = \pm 2$) are elements of finite-dimensional global conformal multiplets \cite{Pasterski:2021fjn, Pasterski:2021dqe, Guevara:2021abz}. The properties of these multiplets are summarized by celestial diamonds\footnote{At $\Delta = -1$ the diamond degenerates to a line.} in the $( \Delta,J)$ plane \cite{Pasterski:2021fjn, Pasterski:2021dqe},  where the left and right corners represent soft modes related by a shadow transform, while the top and bottom corners\footnote{The $\Delta$ axis is taken to be pointing downwards.} represent generalized conformal primaries \cite{Pasterski:2020pdk} the soft modes descend from and to respectively (see Fig. \ref{fig:Diamond2}). Moreover, the bottom corners of the negative helicity soft graviton diamonds can be shown to be primary descendants of $\Delta = 3$ and  $J =-1, 0, 1, 2$ respectively. These coincide precisely with the spin-$s$ operators defined in \eqref{soft-symm} for $s =-1, 0, 1, 2$ or equivalently, the quadratic components $q_s^2$ of the renormalized charges \eqref{Qren}. 

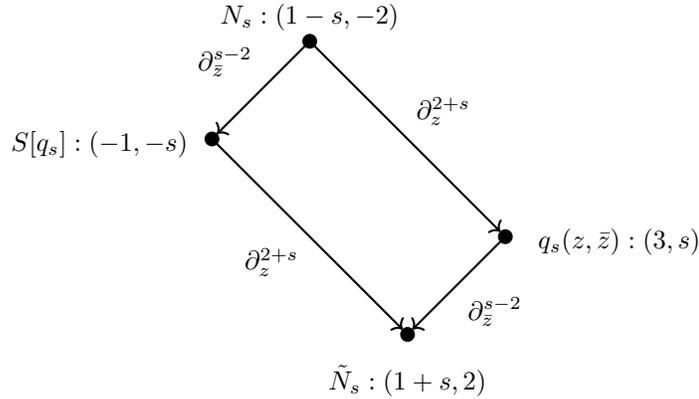
\begin{figure}[h!]
     \centering
       \begin{tikzpicture}[scale=1.3]
\begin{pgfonlayer}{foreground}
\draw[thick](0,2)node[above]{{\footnotesize$N_s:(1-s, -2)$}} ;
\draw[thick,->](0,2)--node[above left]{{\footnotesize$\p_{\bz}^{s-2}$}} (-1+.05,1+.05);
\draw[thick,->](0,2)--node[above right]{{\footnotesize$\p_z^{2 + s}$}} (2-.05,0+.05);
\draw[thick,->] (-1+.05,1-.05)node[left]{{\footnotesize$S[q_s]:(-1, -s)~~$}} --node[below left]{{\footnotesize$\p_z^{2 + s}$}} (1-.05,-1+.05) ;
\draw[thick,->] (2-.05,0-.05)node[right]{{\footnotesize$~~~q_s(z,\bz): (3,s)$}} --node[below right]{{\footnotesize$\p_{\bz}^{s-2}$}} (1.05,-1+.05) ;
\draw[thick](1,-1.2)node[below]{{\footnotesize$ \tilde{N}_{s}:(1+s, 2)$}};
\filldraw[black] (0,2) circle (2pt) ;
\filldraw[black] (2,0) circle (2pt) ;
\filldraw[black] (-1,1) circle (2pt) ;
\filldraw[black] (1,-1) circle (2pt) ;
\end{pgfonlayer}
\end{tikzpicture}
 \caption{Diamond associated with a negative helicity soft graviton of dimension $\Delta = 1 - s$ for $s \geq 3$. The weigth labels are $(\Delta,J)$. Operators connected by long edges have weights  related by $(h, \bar{h}) \leftrightarrow (1 - h, \bar{h})$. Operators connected by short edges have $(h, \bar{h}) \leftrightarrow (h, 1 - \bar{h})$. Diagonally opposite corners are related by $(h, \bar{h}) \leftrightarrow (1 - h, 1 - \bar{h})$.}
    \label{fig:Diamond}
\end{figure}

For $J \geq 3$ similar diamonds can be constructed, however negative (positive) helicity soft graviton modes now lie at the top, while the corresponding charges \eqref{soft-symm} (\eqref{soft-symm-ah}) lie at the right (left) corner. Dimensional analysis reveals that for arbitrary $s$, the weights of opposite corner entries are related by duality $(\Delta, J) \leftrightarrow (2 - \Delta, -J)$, while those of entries  connected by long and short edges are related by $(\Delta, J) \leftrightarrow (1 - J, 1 - \Delta)$ and $(\Delta, J) \leftrightarrow (1 + J,  \Delta-1)$ respectively.  This is summarized in Fig. \ref{fig:Diamond}.

While these relations are suggestive of shadow- and light-transforms respectively, a quick analysis shows that only left and right corners can be mutually non-locally related by shadow transforms. On the other hand, the spin-$s$ charges can be obtained from conformally soft gravitons by simply taking derivatives as in \eqref{soft-symm}, \eqref{soft-symm-ah}.
These features can be understood in terms of the representation theory of complex unimodular groups \cite{Gelfand-Book}. In the case of SL$(2,\mathbb{C})$, the weights $(\Delta,J)$ label representations\footnote{\label{Vrep}
$V_{(\Delta,J)}$ where $\Delta \in \C$ and $ J \in \Z/2$ is the space of analytic functions $\phi(z)$ such that its inversion given by 
$\hat{\phi}(z):= z^{-2h} {\bar{z}}^{-2\bar{h}} \phi(-z^{-1}) $ is also analytic, where $h=(\Delta +J)/2$ and $\bar{h} = (\Delta -J)/2$. This means in particular that $\phi(z)$ admits a Taylor expansion and an asymptotic expansion
\be
\phi_{(h,\bar{h})}(z)\sim  z^{-2h}{\bar{z}}^{-2\bar{h}} \sum_{n,m=0}^{\infty} \frac{\phi_{n,m}}{ z^{n}{\bar{z}}^{m}}
\ee 
when $|z|\to \infty$. These data allow us to construct a smooth function on $\C^2_*$ homogeneous of degree $(-2h,-2\bar{h})$ in $(z_\alpha,\bz_\alpha)$. This function is given by 
\be 
\Phi_{(h,\bar{h})}(z_0,z_1)= z_0^{-2h} \bz_0^{-2\bar{h}} \phi_{(h,\bar{h})} ( z_1/z_0) =(-1)^{2J}
z_1^{-2h} \bz_1^{-2\bar{h}} \hat\phi_{(h,\bar{h})}(- z_0/z_1).
\ee In celestial holography one often works in a more restrictive functional space $V_{h}\otimes V_{\bar{h}}\subset V_{(h+\bar{h},h-\bar{h})}$ in which $z$ and $\bz$ are treated as independent variables. In this functional space $\phi(z)$ also admits expansions of the form
\be 
\phi_{(h,\bar{h})}(z, \bz)\sim z^{-2h} \sum_{n=0}^{\infty} \frac{\bar\phi_h^{n}(\bz)}{ z^{n}}
\sim {\bar{z}}^{-2\bar{h}}\sum_{m=0}^{\infty} \frac{\phi_{\bar{h}}^{m}(z)}{\bz^{m}}\,,
\ee 
where the first expansion is around $z=\infty$ while $\bar{\phi}_h^n$ is assumed to be analytic in $\bz$; similarly the second expansion is around $\bz=\infty$ while  ${\phi}_{\bar{h}}^m$ is analytic. The mode coefficients in these conventions are related to the ones of the ``conformally covariant'' mode expansions by a shift $n \rightarrow n + h, m \rightarrow m + \bar{h}$ for $h, \bar{h} \in \frac{1}{2}\mathbb{Z}$.
} $V_{(\Delta,J)}$ acting on $L^2(\C)$.  These representations are irreducible  unless $\Delta \in \mathbb{Z}$ and either $\Delta > |J|$ or $\Delta \leq -|J|$. It can be shown that in these cases \cite{Gelfand-Book} the representations {admit invariant subspaces and hence} are \emph{reducible} or discrete.

Discrete representations 
 admit decompositions of the form
\be\label{dec1}
V_{(\Delta,J)} = P_{(\Delta,J)}\oplus F_{(\Delta,J)},
\ee 
where $P_{(\Delta,J)}$ is finite dimensional while $F_{(\Delta,J)}= V_{(\Delta,J)}/P_{(\Delta,J)}$ is infinite dimensional.
For negative weights, the discrete representations $P_{(\Delta,J)}$ are simply the space of polynomials of degree $(-\Delta-J,-\Delta+J)$ in $(z,\bz)$. It follows that the maps 
\be
\pa_z^{-\Delta-J+1}&:V_{(\Delta,J)}\to V_{(1-J,1-\Delta)},\cr 
\pa_{\bar{z}}^{-\Delta+J+1}&:V_{(\Delta,J)}\to V_{(1+J,\Delta-1)},
\ee 
annihilate the polynomial subspaces.
These maps therefore
identify the quotient space with the homogeneous space
\be\la{iso}
F_{(\Delta,J)}=V_{(1-J,1-\Delta)}\stackrel{S}{=}V_{(1+J,\Delta-1)}.
\ee 
The last isomorphism is given by the shadow transform denoted by $S[\cdot]$. Finally we have a duality pairing between $F_{(\Delta,J)}$ and $F_{(2-\Delta,-J)}$ given by 
\be
(\phi|\psi)= \int_{\C}\rd^2 z [\pa_z^{-\Delta-J+1}\pa_{\bar{z}}^{-\Delta+J+1} \phi] \psi\,,
\ee 
for for $\psi, \phi \in F_{(\Delta,J)}$.

We see that the celestial diamonds compactly capture this general theory, with the top corners labelled by  discrete representations {of negative weights} and the bottom ones labelled by their duals.
Moreover, the (sub)$^s$-leading soft gravitons $N_s$ are negative discrete when $s\geq 3$ and the corresponding polynomials are  of degree $(s+1,s-3)$. The long  arrows {connecting $N_s$ and $q_s$} in Fig. \ref{fig:Diamond2}, \ref{fig:Diamond} express the isomorphism
 $F_{(1-s,{-}2)}=V_{(3,s)}\stackrel{S}{=} V_{(-1,-s)}$, while $\tilde{N}_s = \pa_z^{s+2}\pa_{\bar{z}}^{s-2} N_s$ is the dual soft graviton.

 Thanks to the analysis of Section \ref{sec:LEE}, we can identify the dual soft graviton with subleading components of $\Psi_0$
 \be
 \tilde{N}_n = \Psi_0^{(n-2)}\,,
 \ee
with the local component corresponding to the image  of the map $\pa_{\bar{z}}^{n-2} $ and the global component in the decomposition \eqref{Psi0} corresponding to the kernel of the map $\pa_{\bar{z}}^{n-2} $.
More precisely, the diamond in Fig. \ref{fig:Diamond} contains two maps $D^{2+s}: V^{-2} \to V^{s}$ which is surjective but contains a kernel 
$K_s= \oplus_{\ell= -2}^{-s+1}  V^{-2}_\ell$ and the second map  $\bar{D}^{s-2}: V^{s} \to V^{2}$ which is injective  but not surjective. It contains a co-kernel ${\tilde K}_s= \oplus_{\ell= 2}^{s-1}  V^{2}_\ell$ which corresponds to the global part of $\Psi_0$.
The fact that the kernel of $D^{2+s}$ is isomorphic to the cokernel of $\bar{D}^{s-2}$ is due to the fact that the shadow transform $S: N_s \to \tilde{N_s}$ is an isomorphism of Lorentz modules.
The dimension of the kernel can be easily evaluated since   $K_s= {\rm ker}[ \bar D^{s-2}]$ is spanned by harmonics $\bar Y^2_{\ell,m}$ of   spin $s - 1 \geq \ell \geq 2$.
This means that 
\be
{\rm dim}\left({\rm ker}[ \bar D^{s-2} ]\right)=\sum_{\ell=2}^{s-1}(2\ell+1)= (s+2)(s-2)\,,
\ee
and this corresponds to the dimesnion of the free parameters in the parametrisation of the global charges  in \eqref{Psi0G}.

\subsection{Soft currents and $w$-currents}\la{sec:wcurr}
In this section we clarify the relationship between the soft graviton and the $w$-current as well as the role of the light transform. 

Soft gravitons {$N_s$} are operators of weights
$(h,\bar{h})=\left( -\frac{1+s}{2}, \frac{3-s}{2}\right).$ As discussed in the previous section, this implies that they fall into discrete $\mathrm{SL}(2,\C)$ representations for $s\geq 3$. Moreover, according to  \eqref{dec1} they admit a decomposition into irreducible components namely $N_s = H_s +\check{N}_s$, where 
$H_s$ is a polynomial in $z$ while $\check{N}_s$ has a Laurent series expansion
\be 
H_s(z,\bz)= \sum_{n=0}^{s+1} z^{n} N_s^{-n}(\bz),
\qquad
\check{N}_s(z,\bz) =\sum_{n=1}^{\infty} \frac{ N_s^{n}(\bz)} {z^n}.
\ee 
On the one hand, since the derivative operator {$\p_z^{s+2}$} annihilates $H_s$, the soft charge is {formally} encoded in the Laurent component as
\be
q_s^1  =\pa_z^{s+2} \check{N}_s({z},\bz)
= \sum_{n=1}^{\infty} \frac{(-1)^{s} N_s^n(\bz)}{z^{(s+2+n)}} \frac{(s+1+n )!}{(n-1)!}.
\ee 
The polynomial component determines, on the other hand, the $w$-current\footnote{ Here we label current by their spin $s$ while in \cite{Himwich:2021dau} they are labelled by the half integer $q= (s+3)/2$. In other words $W_s^{\mathrm{{\scriptscriptstyle here}}} = w^{\frac{s+3}{2}}_{\mathrm{{\scriptscriptstyle there}}}$.} \cite{Guevara:2021abz,  Strominger:2021lvk, Himwich:2021dau}
\be \label{w-currents-def}
W_s(z,\bz):=\sum_{n=0}^{s+1}\frac{(-1)^{(n+s)} N_s^{-n}(\bz)}{z^{(s+2-n)}} n! (s+1-n )!.
\ee

 In \cite{Himwich:2021dau} it is argued that these $w$-currents of   dimension/helicity $(\Delta,J)= (3,s)$ (the same dimension as $q_s^1$) are constructed from light transforms defined as
\footnote{ In \cite{Himwich:2021dau} the light transform for positive helicity graviton is considered.} 
\be 
{\rm \bf L}[O_{(h, \bar{h})}](z, \bz) := \int_{\mathbb{R}} \frac{d {w}}{2\pi i} \frac{1}{(z - {w})^{2 - 2 {h}}} O_{(h, \bar{h})}(w, \bar{z}).
\ee
Such transformations are justified upon analytic continuation to $(2,2)$ signature spacetimes.
Applying this transformation to the negative helicity graviton $G_\Delta^-$ yields a field of dimension $(3,1-\Delta)$.
Nevertheless, it turns out that in the limit $\Delta \to 1-s$, singularities arise that have not been properly accounted for in previous discussions. In particular, light transforms of fields of negative {weights} are singular,
meaning that $W_s$ cannot be simply characterized as the limit $\epsilon\to 0$ of the the light transform of $G_{1-s+\epsilon}^{{-}}$.  What we find instead is that 
\be \label{Lightt}
(-1)^{(s+3)} \Gamma(s+3) {\rm \bf L}\left[G^{{-}}_{ 1-s + \epsilon} \right](z, \bz) = \frac{q_s^1(z,\bz)}{\epsilon} + W_s(z,\bz) + o(\epsilon).
\ee 
In other words, the $w$-current appears as the renormalized light transform
\be \la{LT}
W_s &=  \lim_{\epsilon \rightarrow 0}\left( 
(-1)^{(s+3)} \Gamma(s+3){\rm \bf L}\left[G_{ 1-s + \epsilon}   \right]-\frac{q_s^1}{\epsilon}\right) := {\rm \bf L}\left[N_s\right].
\ee 
The proof of this statement is given in Appendix \ref{App:wcurr}. 
We see that the $w$-current and the soft {charge} correspond to different mode  {projections} of the soft graviton in the limit $\Delta \to 1-s$. In particular, the soft charge $q_s^1$ arises from the singular component of ${\rm \bf L}[G_{1-s +\epsilon}]$, while the $w$-current is extracted from the regular component of ${\rm \bf L}[G_{1-s +\epsilon}]$. Moreover, as shown in \cite{Strominger:2021lvk}, this current is the image of the  polynomial soft graviton of degree $(s+1)$ in $z$.

\subsection{$w_{1 + \infty}$ structure from charge recursion}\la{sec:recursion}

Despite these distinctions, in this section we demonstrate an intriguing relation between OPEs involving the spin-$s$ charges and OPEs involving the $w$-currents constructed from the light-transform in \cite{Strominger:2021lvk, Himwich:2021dau}.

We start with the delta-function identity
\be 
\p_x^n \delta(x) = \frac{(-1)^n n!}{x^n} \delta(x)
\ee
or equivalently
\be \la{pd}
\p_{z_1}^{s - n} \delta^{(2)}(z_{12}) = \frac{(-1)^{s - n}(s - n)!}{z_{12}^{s - n}} \delta^{(2)}(z_{12}),
\ee
and re-express the symmetry action  \eqref{rec} of the soft graviton  as
\be 
\label{holo-ope}
\begin{split}
q_s^{1}(z_1) G^{\pm}_{\Delta_2}(z_2, \bz_2) &\sim  \f\k{8i} \sum_{n = 0}^s  \frac{(n + 1) (2h_{2\pm})_{s - n}}{z_{12}^{s - n}} \delta^{(2)}(z_{12}) \p_{z_2}^n G^{\pm}_{\Delta_2 + 1 - s}(z_2, \bz_2).
\end{split}
\ee
We recall that the identities derived from same-helicity OPEs hold in general, while the ones obtained from opposite helicity OPEs are only valid in holomorphic and antiholomorphic collinear limits respectively as discussed in Section \ref{ss:celestial}.

On the other hand, in \cite{Himwich:2021dau} $w$-currents of the same spin were shown to obey the following OPEs \cite{Himwich:2021dau}\footnote{To compare with \cite{Strominger:2021lvk, Himwich:2021dau} one needs to set $2q=s+3$. Moreover, our normalization of the $w$-current differs by a factor $\frac{\kappa^2 i}{8\pi}$ from the one employed there.}
\be 
\label{w-current}
W_s(z,\bz) {O}_{(h, \bar{h})}(0,0) \sim -\frac{{\kappa^2}}{{16\pi i} \bz} \sum_{n = 0}^{s} \frac{(n + 1) \Gamma(2h + 1)}{\Gamma(2 h + 1 - s + n)} z^{n - s-1 } \p^n 
O_{ \left( (2h+1- s)/{2},  (2\bar{h} +1 -s)/{2}\right) }(0,0).
\ee
Simplifying the ratio of Gamma functions and letting $O$ be a graviton, \eqref{w-current} reduces to
\be 
\label{w-simple}
W_s(z,\bz)  G^{\pm}_{\Delta}(0,0) \sim -\frac{{\kappa^2}}{{16\pi i} z\bz} \sum_{n = 0}^{s} \frac{(n + 1) (2 h_{\pm})_{s - n}}{ z^{s - n }} {\p}^n G_{\Delta - s + 1}^{\pm}(0,0).
\ee

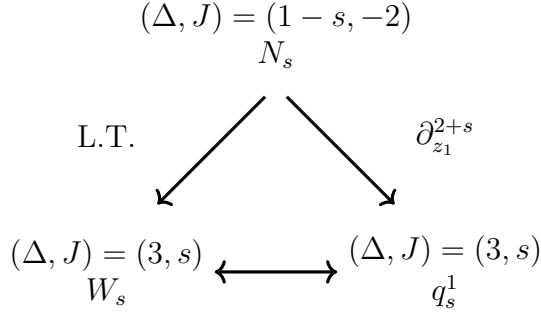
\begin{figure}[ht]
    \centering
\begin{tikzpicture}[scale = 4.5]
\begin{pgfonlayer}{foreground}
\node[] (int0) at (1.3,-0.2) {\shortstack[position] {$(\Delta,J) = (1 - s, -2) $ \\ $ N_s $}};
\node[align=center] (int1) at    +(0.8,-.9) {\shortstack[position] {$(\Delta,J) = (3, s) $ \\ $ W_s $}};
\node[] (int2) at   +(1.8,-.9) {\shortstack[position] {$(\Delta,J) =
(3, s) $ \\ $ q_s^1$
}};
\node[] (int) at   (1.3,-.35) {};
\draw[->, very thick, color=black] (int) -- +(-.35,-.35);
\draw[->, very thick, color=black] (int) -- +(.35,-.35);
\draw[<->, very thick, color=black] (int1) -- (int2);
\node[] (int) at   (0.8,-.5) {L.T.};
\node[] (int) at   (1.8,-.5) {$\p_{z_1}^{2 +s}$};
\end{pgfonlayer}
\end{tikzpicture}
   \caption{There are two maps from a soft graviton $N_s$ with $(\Delta, J) = (1 - s, -2)$ to an operator of $(\Delta, J) = (3, s):$ the light transform defined in \eqref{LT} and the action of $s + 2$ derivatives $\p_{z_1}^{2 + s}$. The resulting operators have the \textit{same} OPE with massless celestial operators upon trading $2\pi\delta^{(2)}(z)$ for $1/(z\bz)$.}
    \label{fig:GD}
\end{figure}
Remarkably, comparing \eqref{holo-ope} and \eqref{w-simple}, we see that while they differ in their singularity structures, their OPE data are identical after the replacement of $1/z\bar{z}$ with the contact term $2\pi\delta^{(2)}(z)$. We summarize this in Figure \ref{fig:GD}. 
The $w$-currents \eqref{w-currents-def} were shown to generate a $w_{1 + \infty}$ algebra in \cite{Guevara:2021abz,Strominger:2021lvk,Himwich:2021dau}, we take this as strong evidence that the higher spin charges \eqref{Q2} similarly generate a $w_{1 + \infty}$ symmetry. Remarkably, in the next section we show that this indeed turns out to be true at the linear order in the algebra. We leave an explicit check to higher orders, as well as the classical corrections arising from the collinear contributions \eqref{charge-kexp} for $k \geq 3$ to future studies.

\subsection{Charge bracket}\la{sec:CB}

We conclude our analysis by showing that the higher spin charge aspects \eqref{Qren},  \eqref{Qren2} provide a realization of the $w_{1+\infty}$ algebra at linear order. We present here the main steps of the calculation of the linear part of their Poisson bracket, namely we compute
\be\la{b12}
\{q_s(z),  q_{s'}(z')\}^1=\{q^2_s(z),  q^1_{s'}(z')\}+\{q^1_s(z),  q^2_{s'}(z')\}\,.
\ee
The details are deferred to Appendix \ref{app:CB}.
To compute the algebra above we need to take and extra derivative $D_{z'}^{s'+2}$ of \eqref{q2Ns}. Starting with the first bracket in \eqref{b12}, this gives 
\be
\{q^2_s(z),  q^1_{s'}(z')\}&=
\frac{\kappa^2}{8} \sum_{n = 0}^s(n+1) 
\left(\begin{matrix}
s'+s-n\\
s'
\end{matrix}\right)
D_{z'}^{s'+2}\left( D_{z'}^n  N_{s'+s-1}(z')D_{z}^{s - n} \delta(z, z')\right)\cr
&=\frac{\kappa^2}{8}  \sum_{p=0}^{s+s'+2}G(s,s',p)
\left( D_{z'}^{1-p}  q^1_{s'+s-1}(z')\right) D_{z}^{p} \delta(z, z')\,,\la{q12a}
\ee
where
\be
G(s,s',p):= \sum_{n = {\rm max}[0,p-s'-2]}^{{\rm min}[p,s]} (-)^{p+n}(s-n+1)
\left(\begin{matrix}
s'+n\\
s'
\end{matrix}\right)
\left(\begin{matrix}
s'+2\\
p-n
\end{matrix}\right)\,.
\ee
In  Appendix \ref{app:CB} we show that 
\be \la{G11}
G(s,s',p)=0 \,,\quad  {\rm when}~2\leq p\leq s+1\,,
\ee
while 
\be
G(s,s',0)= 
1+s\qquad {\rm and}\qquad  G(s,s',1)=
-(2+s+s').
\ee
Moreover, we find that when  $p\geq s+2 $, 
\be \la{G3}
G(s,s',p) = \f{(-)^{p+s}(s+s'+2)!}{(p-s-2)!(s+s'+2-p)! s!}
\f1{p(p-1)}.
\ee 
Therefore  we obtain the final expression
\be\la{q2q1}
\{q^2_s(z),  q^1_{s'}(z')\}
&=\frac{\kappa^2}{8}\bigg[ 
 (s+1)D_{z'} \left( q^1_{s'+s-1}(z')  \delta(z, z') \right)
 -(s'+1) q^1_{s'+s-1}(z') D_{z} \delta(z, z')\\
 &+\sum_{p=s+2}^{s+s'+2}G(s,s',p)
\left( D_{z'}^{1-p}  q^1_{s'+s-1}(z')\right) D_{z}^{p} \delta(z, z')\bigg]\,.\nonumber
\ee

The second bracket in 
\eqref{b12} is obtained after the exchange $s\leftrightarrow s', z\leftrightarrow z'$ as
\be
\{q^1_s(z),  q^2_{s'}(z')\}
&
=
-\frac{\kappa^2}{8}\sum_{p=0}^{s+s'+2}
G(s',s,p) \left( D_{z}^{1-p}  q^1_{s'+s-1}(z)\right) D_{z'}^{p} \delta(z, z')\,.
\ee
An analogous split of the sums  allows us to write the RHS as 
\be
\{q^1_s(z),  q^2_{s'}(z')\}&=-\frac{\kappa^2}{8}  \sum_{p=0}^{s+s'+2}  
G(s',s,p)
\left( D_{z}^{1-p}  q^1_{s'+s-1}(z) \right)D_{z'}^{p} \delta(z, z')\cr
&=-\frac{\kappa^2}{8} \bigg[
(s'+1)D_{z}(  q^1_{s'+s-1}(z)  \delta(z, z'))
 -(s+1) q^1_{s'+s-1}(z) D_{z'} \delta(z, z')\cr
 &+\sum_{p=2}^{s+s'+2}
G(s',s,p) \left( D_{z}^{1-p}  q^1_{s'+s-1}(z)\right) D_{z'}^{p} \delta(z, z') \bigg]\,.\label{breverse}
\ee
The last sum above can be recast as
\be
&\sum_{p=2}^{s+s'+2} G(s',s,p)
\left( D_{z}^{1-p}  q^1_{s'+s-1}(z)\right) D_{z'}^{p} \delta(z, z')\cr
&= \sum_{m=0}^{s+s'+2} \left(\sum_{p={\rm max}[m,2]}^{s+s'+2} (-)^m  \left(\begin{matrix} p\\ m\end{matrix}\right)G(s',s,p) \right)
\left( D_{z'}^{1-m}  q^1_{s'+s-1}(z')\right) D_{z}^{m}\delta(z, z')\,,
\la{b21}
\ee
where we have used Leibniz rule to exchange the $z$ and $z'$ derivatives and exchanged the sums.
It can be shown that the terms   in \eqref{b21} cancel exactly the RHS  of \eqref{q2q1} upon adding up the two brackets in \eqref{b12}.  This is due to the property (see proof in Appendix \ref{app:CB})
\be
&\sum_{p=m}^{s+s'+2} (-)^m 
\left(\begin{matrix}
p\\
m
\end{matrix}\right)
G(s',s,p)=G(s,s',m)\qquad{\rm for}\quad 0\leq m\leq  s+s'+2\,.
\ee
We are thus left only with the local  terms $p=0,1$ in \eqref{breverse} and the final result is
\be
\{q_s(z),  q_{s'}(z')\}^1
&=\frac{\kappa^2}{8}\left[ 
-\left(s'+1\right)  q^1_{s'+s-1}(z') D_{z} \delta(z, z')
+ (s+1) 
q^1_{s'+s-1}(z) D_{z'} \delta(z, z')
\right]\,,\la{WTF}
\ee
corresponding to a $w_{1+\infty}$ algebra. 
In terms of the higher spin charges \eqref{Qs} the algebra \eqref{WTF} takes the form
\begin{empheq}[box=\fbox]{align} 
\{Q_s(\tau),  Q_{s'}(\tau')\}^1=
\left(s'+1\right) Q^1_{s'+s-1}(\tau' D\tau)
- (s+1) 
 Q^1_{s'+s-1}(\tau D\tau')\,.\la{WTF2}
\end{empheq}
It is important to note that  the transformation parameters $\tau_s(z,\bz)$  belong to the SL$(2,\mathbb{C})$ representations $ V_{(-1, -s)}$ with weights $(h,\bar{h})= (-\tfrac{s+1}{2}, \tfrac{s-1}{2})$. Similarly   $\tau_{s'}'(z',\bz') \in V_{(-1, -s')}$.
This means that we can perform an asymptotic expansion (see footnote \ref{Vrep})
\be
\tau(z,\bar{z}) &=\sum_{m\geq 0} z^{ {s + 1} - m} \tau_s^m(\bar{z})\,,
\ee
where $\tau_{s}^m (\bar{z})$ 
also admits an asymptotic expansion $ \tau_{s}^m (\bar{z}) = \sum_{n\geq 0} \bar{z}^{1-s-n} \tau_s^{m,n}$ 
and similarly for ${\tau'}_{s'}^{m'}$.
We denote the charge associated with the mode function $\tau_s^{m,n}(z,\bar{z}):=z^{ {s + 1} - m} \bar{z}^{1-s-n} $ by $Q^s_{m,n}$. 
From \eqref{WTF2} we then find the loop  algebra $Lw_{1+\infty}$  
\be 
\begin{split}
[Q_{m,n}^s, Q_{m',n'}^{s'}]
&= i\left[m(1 + s') - m' (1 + s) \right]Q^{s + s' - 1}_{m + m'-1, n+n'}\,,
\end{split}
\ee
where $m,n\in \mathbb{N}$.
The wedge subalgebra $WLw_{1+\infty}\subset Lw_{1+\infty}$  is obtained by restricting to parameters $\tau$ to be polynomials of degree $s+1$ in $ z$. This amounts to the restriction  $m \leq s+1$. This wedge subalgebra is described in \cite{Adamo:2021lrv} as the symmetry of the twistor formulation of self-dual gravity.\footnote{ To compare with  \cite{Adamo:2021lrv}, we need to use that  $q^s_{m}(\bar{z})= w^{\frac{s+3}{2}}_{\frac{s+1}{2} +m}(\bar{z})$. In particular, $m_{\rm {\scriptscriptstyle here}} = \frac{s+1}{2} +m_{\rm {\scriptscriptstyle there}}$.}   We see here that there is no need from the canonical analysis to make this restriction.

\section{Conclusions}

Motivated by the analysis of the gravitational phase space at null infinity in \cite{Freidel:2021qpz, Freidel:2021dfs}, we have proposed a set of evolution equations for  higher spin-$s$ charges.
We conjectured that this extension encodes a truncation of the asymptotic gravitational dynamics at subleading orders in a large-$r$ expansion. After explaining how these charges should appear in the expansion of the Weyl scalar encoding incoming radiation data (see Eq. \eqref{Psi0}) and explicitly proving our conjecture in the case $s=3$, we have investigated the implications of \eqref{highQ}  for the symmetry content of gravity.
The higher spin evolution equations define, after a regularization procedure, 
a representation of the  higher spin-$s$ charges on the gravity phase space.   This representation generalizes the leading, subleading and sub-subleading Einstein's evolution equations at $\scri$ to the case $s\geq 3$. 

Upon introduction of a proper renormalization of the charges, we {computed the} action of their quadratic contribution on the asymptotic shear \eqref{bulk-s-cfin}, \eqref{bulk-s-cbfin}. On the one side, this result has allowed us to obtain the pseudo-vector fields \eqref{dtauC}, \eqref{dtaubC} associated to the transformations generated by the higher spin charges and to derive an infinite tower of soft graviton theorems \eqref{STs} (truncated at quadratic order, that is neglecting collinear terms) induced by their conservation laws.  This generalizes our previous results obtained in  \cite{ Freidel:2021dfs} to all $s$. On the other side, we have shown that this action  reproduces exactly the OPE \eqref{qNOPE} between  soft charges and soft graviton operators obtained through  celestial holography techniques. Moreover, we have elucidated how the same OPE structure is reproduced when 
replacing the soft charges with the  $w$-currents of the same spin introduced in \cite{Himwich:2021dau}.  To shed some light into this interesting feature, we clarified how the light transform of the soft graviton \eqref{Lightt} contains both a singular component given by the soft current, corresponding to the local charge, and a regular one given by the $w$-current and corresponding to the global charge. 

As the $w$-currents have been shown to generate an infinite  higher spin celestial symmetry algebra \cite{Strominger:2021lvk, Himwich:2021dau}, we have completed our canonical analysis by proving to linear order that   the loop algebra $Lw_{1+\infty}$  has a canonical realisation in the gravitational phase space in terms of the  Poisson bracket of the higher spin charges \eqref{WTF2}. This provides evidence of a spacetime interpretation of such a new infinite dimensional symmetry beyond the self-dual gravitational sector. 

To fully understand the role of the   $Lw_{1+\infty}$ algebra in gravity we need to extend our analysis in two intertwined directions.
On the one hand, the relevance of the recursion relation \eqref{dotQss} in encoding the expression of the vacuum Einstein's equations at subleading orders in a large-$r$ expansion needs to be firmly established beyond the $s\leq 3$ case. On the other hand, one needs to investigate whether the $Lw_{1+\infty}$ algebra structure survives the inclusion of the non-linear corrections, which include quadratic same helicity contributions and  higher order contributions. More precisely,  we have pointed out in Section 
\ref{subs:hssg} that the recursion relation for $\cQ_s$ \eqref{dotQss} acquires corrections purely quadratic  in the same helicity fields
$\bar C$ in order to correctly reproduce the vacuum Einstein's equations for
spin $s\geq 4$. This extra quadratic corrections do not affect the same helicity  linear bracket \eqref{WTF2} $\{q_s,q_{s'}\}^1$ and $\{\bar{q}_s,\bar{q}_{s'}\}^1$.
However, the presence of these quadratic corrections will affect the mixed helicity charge bracket $\{{q}_s,\bar{q}_{s'}\}$ already at linear order. Such corrections are in fact expected also from the point of view of the celestial OPE calculation, as recalled  at the end of Section \ref{ss:celestial}. 
Moreover, one needs to show explicitly that the loop algebra $Lw_{1+\infty}$ for the same helicity charges is  valid at quadratic order as well. Evidence that this is the case has already been given in \cite{Himwich:2021dau} but a direct derivation is still needed from our perspective.

Our analysis suggests that contact terms should play an important role in celestial conformal field theories. It would be interesting to revisit the analyses relying on celestial OPE expansions carefully accounting for potential contact terms.
Relatedly, one can wonder whether the $Lw_{1+\infty}$ survives or not the introduction of the higher order colinear contributions. 
In fact, demanding that the symmetry is preserved through the introduction of  non-linearities would result in powerful constraints on the (sub)$^s$-leading dynamics.

Despite these open issues, what we find remarkable is the so far perfect match and the emergence of a precise dictionary between the two side of the asymptotic symmetry story, namely the celestial CFT  description of the S-matrix scattering amplitudes and the structure of Einstein's equations expanded around null infinity.


\section*{Acknowledgements}

We thank Sabrina Pasterski for a discussion about the celestial diamond.
Research at Perimeter Institute is supported in part by the Government of Canada through the Department of Innovation, Science and Economic Development Canada and by the Province of Ontario through the Ministry of Colleges and Universities. This project has received funding from the European Union's Horizon 2020 research and innovation programme under the Marie Sklodowska-Curie grant agreement No. 841923. A.R. is additionally supported by the Stephen Hawking fellowship.

\appendix

\section{Linearized Einstein equations}
\label{app:edth}
In this appendix we relate the conventions in this paper with those of \cite{Newman:1968uj}. We establish a relation between $D$ and the edth operator by considering the action 
\be 
\begin{split}
D O_s \equiv m^A m^{A_1} \cdots m^{A_s} D_A O_{A_1 \cdots A_s} &= (D_m - s m^A D_m \bar{m}_A) O_s\\
&= (P \p_z + s (\p_z P)) O_s = P^{1 - s} \p_z \left(P^{s} O_s \right)\,,
\end{split}
\ee
where $D_m = m^A D_A$ and we used 
\be
m^A D_m \bar{m}_A= m^A m^B (\pa_B \bar{m}_A -\Gamma_{AB}^C \bar{m}_C)= P^2 \pa_z P^{-1} - \Gamma_{zz}^{\bar{z}} P  = -\pa_z P.
\ee 
Similarly, it can be shown that
\be 
\bar{D} O_s = P^{1 + s} \p_{\bz}(P^{-s} O_s).
\ee
Since the action of the edth operator can be written as
\cite{Goldberg, eastwood_tod_1982,  doi:10.1063/1.528587}
\be
\eth \eta_s =\sqrt{2}P^{1-s}\left[\p_z (P^{s} \eta_s)\right]\,,
\quad 
\bar \eth \eta_s =\sqrt{2}P^{1+s}\left[\p_\bz (P^{-s} \eta_s)\right]\,,
\ee
we thus recover the relations
\be
\eth= \sqrt{2} D\,,\quad \bar\eth=\sqrt{2} \bar{D}\,.
\ee

We now show how to determine the global solution \eqref{eq:gl-sol}, \eqref{eq:gl-alpha}. 
Starting with the Ansatz \eqref{eq:gl-sol}, on the one hand we have that 
\be
\pa_u\Psi_{G0}^{(n+1)}(u) =\sum_{k=0}^{n-1} (n-k) \alpha_{n}^k   G_{k}\, u^{n-1-k},
\ee
while on the other hand, using that  $G_k \in V^2_{2+k}$ where $V^s_\ell$ is the module of spin $s$ and angular momentum $\ell$,
\bea
-\f1{  (n+1)} \left(\bar D D +\frac12 n(n+5) \right) \Psi_{G0}^{(n)}(u) &=&
- \sum_{k=0}^{n-1} \frac{\left(n(n+5)- k(k+5) \right)}{2(n+1)} \alpha_{n-1}^k   G_{k}\, u^{n-1-k} \cr
&=&- \sum_{k=0}^{n-1}  \frac{(n-k) (n+k+5)}{  2 (n+1)} \alpha_{n-1}^k   G_{k}\, u^{n-1-k}.
\eea
In the last equality we used that
\be
n(n+5)- k(k+5) = (n-k)(n+k+5).
\ee
The constraint equations \eqref{dP0n} then imply the recursion relation
\be 
\alpha_n^k = -\frac{n + k + 5}{2(n + 1)} \alpha_{n - 1}^k,
\ee
which subject to the boundary condition $G_n = \left[\Psi_{G0}^{(n + 1)} \right]_{l = 2 + n}$ (ie. $\alpha_n^n = 1$) yields
\be 
\alpha_{n}^k = (- 2)^{k - n} \frac{(k + 1)! (n + k + 5)!}{(n + 1)!(2k + 5)!}.
\ee

\section{Soft graviton bracket}\la{App:SGbra}

In this appendix, we spell out in detail the computation of the bracket of a
positive helicity spin-$s'$ soft graviton operator with a spin-$s$ charge. Since positive helicity soft gravitons commute with the linear component of the charge \eqref{Qren2}, we only need to consider the bracket with the  quadratic component. By means of \eqref{bulk-s-cfin}, we have
\be
\{q^2_s(z), \bar N_{s'}(z')\}&=\f{(-1)^{s'+1}}2\f1{s'!}\int_{-\infty}^\infty \rd u\, u^{s'} \{q^2_s(z),\hat {\bar N}(u,z')\}\cr
&= \frac{\kappa^2}{8} \sum_{n = 0}^s \f{(-1)^{s'+s+1}}2 \int_{-\infty}^\infty \rd u\,
\frac{(-1)^{ n} ( n+1)}{s'!(s-n)!}\cr
&\times u^{s'} \p_u(\Delta + 2)_{s - n}(\p_{u}^{-1})^{s - 1} D_{z'}^n C (u, z') D_{z}^{s - n} \delta(z, z')\cr
&=\frac{\kappa^2}{8} \sum_{n = 0}^s \f{(-1)^{s'+s+1}}2 \int_{-\infty}^\infty \rd u\,
\frac{(-1)^{ n} ( n+1)}{s'!(s-n)!}(\Delta -s'+3)_{s - n}\cr
& \times u^{s'} (\p_{u}^{-1})^{s - 1} D_{z'}^n {\hat {\bar{N}}} (u, z') D_{z}^{s - n} \delta(z, z')\cr
&=\frac{\kappa^2}{8} \sum_{n = 0}^s \f{(-1)^{s'+s+1}}2 \int_{-\infty}^\infty \rd u\,
\frac{(-1)^{ n} ( n+1)}{s'!(s-n)!}(\Delta -s'+3)_{s - n}\cr
& \times u^{s'+s-1} u^{-s + 1}\p_{u}^{-s + 1} D_{z'}^n {\hat {\bar{N}}} (u, z') D_{z}^{s - n} \delta(z, z')\cr
&=\frac{\kappa^2}{8} \sum_{n = 0}^s\f{(-1)^{s'+s+1}}2  \int_{-\infty}^\infty \rd u\,
\frac{(-1)^{ n} ( n+1)}{s'!(s-n)!}\f{(\Delta -s'+3)_{s - n}}{ (\Delta-s'-1)_{s - 1}} \cr
& \times u^{s'+s-1} D_{z'}^n {\hat {\bar{N}}}(u, z') D_{z}^{s - n} \delta(z, z')\,,
\ee
where
we used the relations
\be
&u^n\p_u^n=(\Delta-1)_n\,,\qquad \p_u^nu^n=(\Delta+n-1)_n\,,\qquad
u^{-n}\p_u^{-n}=(\Delta+n-1)^{-1}_n\,,\cr
& \p_u (\Delta+\alpha)_n= (\Delta+\alpha+1)_n \p_u\,,\qquad
\p_u^{-1} (\Delta+\alpha)_n= (\Delta+\alpha-1)_n \p_u^{-1}\,,\cr 
& u(\Delta+\alpha)_n= (\Delta+\alpha-1)_n u
\,,\qquad
u(\Delta+n-1)^{-1}_n=(\Delta+n-2)^{-1}_n u\,,
\la{uDelta}
\ee
valid $\forall~ n\geq 0\,,\alpha\in \Z$. We now notice that the operator $\Delta=\p_u u$ and any analytic function of it integrate to zero, given our choice of boundary conditions \eqref{Nbc}. More precisely in order for the charge $q_{s+s'-1}$ to be defined we need to demand that  $\hat{N}=O(u^{s+s'-1-\epsilon})$. This means that we can write the bracket in the final form
\be
\{q^2_s(z), \bar N_{s'}(z')\}&=-\frac{\kappa^2}{8} \sum_{n = 0}^s \f{(-1)^{s'+s+1}}{2(s' +s -1)!}( n+1)\f{(s'+s-n-4)!}{ (s-n)! (s'-4)!}\cr&\times 
\int_{-\infty}^\infty \rd u\,u^{s'+s-1} D_{z'}^n \hat {\bar N} (u, z') D_{z}^{s - n} \delta(z, z')\cr
&=\frac{\kappa^2}{8} \sum_{n = 0}^s(n+1) 
\left(\begin{matrix}
s'+s-n-4\\
s'-4
\end{matrix}\right)
D_{z'}^n \bar N_{s'+s-1}(z')D_{z}^{s - n} \delta(z, z')\,.
\ee

A similar calculation for the negative helicity spin-$s'$ soft graviton operator, by means of \eqref{bulk-s-cbfin}, yields
\be\la{q2N}
\{q^2_s(z),  N_{s'}(z')\}&=
\frac{\kappa^2}{8} \sum_{n = 0}^s(n+1) 
\left(\begin{matrix}
s'+s-n\\
s'
\end{matrix}\right)
D_{z'}^n  N_{s'+s-1}(z')D_{z}^{s - n} \delta(z, z')\,.
\ee

\section{Pseudo-vectors}\label{App:pseudo}

We provide some technical details of the spin-$p$ pseudo-vector action on the shear presented in Section \ref{sec:pseudo}.
Using the relation \eqref{uDelta} we get that 
\be
\frac{(\Delta + 2)_{s - n}}{(s-n)!} \pa_u^3 &=
\pa_u^3 \frac{(\Delta -1)_{s - n}}{(s-n)!}
= \pa_u^3\frac{ u^{s-n}}{(s-n)!}\pa_u^{s-n}\cr
&=\sum_{k=0}^{\mathrm{min}[3, s-n]} \left(\begin{matrix}
3\\
k
\end{matrix}\right)  \frac{u^{s-n-k}}{(s-n-k)!}\pa_u^{s-n +3-k}\,.
\ee
Similarly, we evaluate 
\be
\frac{(\Delta - 2)_{s - n}}{(s-n)!} &= 
\pa_u^{-1} \frac{(\Delta  - 1)_{s - n}}{(s-n)!} \pa_u
= \pa_u^{-1} \frac{u^{s-n}\pa_u^{s-n}}{(s-n)!} \pa_u \cr &=
\sum_{k=0}^{s-n}(-1)^k \frac{u^{s-n-k} \pa_u^{s-n-k}}{(s-n-k)!}\,.
\ee 
Therefore 
\be 
\{Q_s^{2}(\tau), \bar{C}(u', z')\} &= \sum_{n = 0}^s 
\frac{(n+1)(\Delta - 2)_{s - n}}{(s-n)!}  (D_{z}^{s - n}\tau_s)    D_{z'}^n \p_{u'}^{1-s} \bar{C}(u', z')\cr
& = \sum_{n = 0}^s 
\sum_{k=0}^{ s-n} (-1)^k
\frac{(n+1)u'^{s-n-k}}{(s-n-k)!}  (D_{z}^{s - n}\tau_s)    D_z^n \pa_{u'}^{1-n-k} \bar{C}(u', z') \cr
&=\sum_{p = 0}^s \frac{u'^{s-p}}{(s-p)!} \sum_{k=0}^{p} 
  (-1)^k (p+1-k)
 (D_{z}^{s - p +k}\tau_s)    [ D^{p-k}\pa_{u'}^{1-p} \bar{C}(u', z')] \cr
 &=\sum_{p = 0}^s \frac{u'^{s-p}}{(s-p)!} \delta^p_{D^{s - p}\tau_s} \bar{C}(u', z')   .
\ee

\section{$\bar C$ bracket}\la{App:barC}

By means of the bracket 
\be
\{C(u,z), \bar C(u', z')\}=-\f\k2 \theta(u'-u) \delta(z,z')\,,
\ee
we have
\be
\{\hat q_s^{2}(u,z), \bar C(u', z')\}
&= \frac14\sum_{n=0}^{s}\sum_{\ell=0}^n \frac{(-u)^{s-n}}{(s-n)!} (\ell+1) 
(\p_u^{-1})^{n-\ell+1}  D^{s-\ell}\left[\{C(u,z), \bar C(u', z')\}  (\p_u^{-1} D)^{\ell} \cN (u,z) \right]\cr
&=-\f\k8 \sum_{n=0}^{s}\sum_{\ell=0}^n \frac{(-u)^{s-n}}{(s-n)!} (\ell+1) 
(\p_u^{-1})^{n-\ell+1}  D^{s-\ell}\left[\theta(u'-u) \delta(z,z')  (\p_u^{-1} D)^{\ell} \cN (u,z) \right]\cr
&=-\f\k8 \sum_{n=0}^{s}\sum_{\ell=0}^n \frac{(-u)^{s-n}}{(s-n)!} (\ell+1) \cr
&\times \p_{u'}^{-1}  \left[(\p_u^{-1})^{n-\ell+1}  \d(u'-u)D_z^{s-\ell} \delta(z,z')  (\p_u'^{-1})^{\ell-2}  D_{z'}^{\ell}\bar C (u',z') \right]\cr
&=\f\k8
 \sum_{n=0}^{s}\sum_{\ell=0}^n \frac{(-u)^{s-n}}{(s-n)!} (\ell+1) \cr
&\times \p_{u'}^{-1}  \left[ \f{(u-u')^{n-\ell}}{(n-\ell)!}D_z^{s-\ell} \delta(z,z')  (\p_u'^{-1})^{\ell-2}  D_{z'}^{\ell}\bar C (u',z') \right]\,,
\ee
where we have used \eqref{id}. We can now switch sums and evaluate  $\sum_{n=\ell}^s$ first. This step makes it explicit that bracket is well defined in the limit $u\to-\infty$ and 
the renormalized charge  yields
\be
\{q_s^{2}(z), \bar C(u', z')\}
&=\f\k8\sum_{n=0}^s (n+1) \p_{u'}^{-1} \left[ \f{(-u')^{s-n}}{(s-n)!} D_z^{s-n}\delta(z,z') (\p_{u'}^{-1} )^{n-2}D^n_{z'} \bar C (u',z') \right]\cr
&=\f\k8\sum_{n=0}^s (-)^{s-n} \f{(n+1)}{(s-n)!}  (\Delta -2)_{s-n} D_z^{s-n}\delta(z,z') (\p_{u'}^{-1} )^{s-1}D^n_{z'} \bar C (u',z')\,,
\ee
where in the last passage we have used again the generalized Leibniz rule \eqref{Leibn2}.

\section{Charge bracket}\la{app:CB}

The bracket \eqref{q2N} allows us to compute the linear charge algebra
\be
\{q^2_s(z),  q^1_{s'}(z')\}
&=\frac{\kappa^2}{8} \sum_{n = 0}^s(s-n+1) 
\left(\begin{matrix}
s'+n\\
s'
\end{matrix}\right)
D_{z'}^{s'+2}\left( D_{z'}^{s-n}  N_{s'+s-1}(z')D_{z}^n \delta(z, z')\right)\cr
&=\frac{\kappa^2}{8} \sum_{n = 0}^s \sum_{m=0}^{s'+2}(s-n+1)
\left(\begin{matrix}
s'+n\\
s'
\end{matrix}\right)
\left(\begin{matrix}
s'+2\\
m
\end{matrix}\right)\cr
&\times
\left( D_{z'}^{s'+s-n-m+2}  N_{s'+s-1}(z')D_{z'}^m D_{z}^n \delta(z, z')\right)\cr
&=\frac{\kappa^2}{8} \sum_{n = 0}^s \sum_{m=0}^{s'+2}(-)^{m}(s-n+1)
\left(\begin{matrix}
s'+n\\
s'
\end{matrix}\right)
\left(\begin{matrix}
s'+2\\
m
\end{matrix}\right)\cr
&\times
\left( D_{z'}^{1-n-m}  q^1_{s'+s-1}(z') D_{z}^{n+m} \delta(z, z')\right)\cr
&=\frac{\kappa^2}{8}  \sum_{p=0}^{s+s'+2} \sum_{n = {\rm max}[0,p-s'-2]}^{{\rm min}[p,s]} (-)^{p+n}(s-n+1)
\left(\begin{matrix}
s'+n\\
s'
\end{matrix}\right)
\left(\begin{matrix}
s'+2\\
p-n
\end{matrix}\right)\cr
&\times
\left( D_{z'}^{1-p}  q^1_{s'+s-1}(z') D_{z}^{p} \delta(z, z')\right)\,\cr 
&=\frac{\kappa^2}{8}  \sum_{p=0}^{s+s'+2}
G(s,s',p) \left( D_{z'}^{1-p}  q^1_{s'+s-1}(z') D_{z}^{p} \delta(z, z')\right),\la{qq}
\ee
where we defined
\be
G(s,s',p):=\sum_{n={\rm max}[0,p-s'-2]}^{{\rm{min}}[s,p]} (-)^{p+n}(s-n+1)
\left(\begin{matrix}
s'+n\\
s'
\end{matrix}\right)
\left(\begin{matrix}
s'+2\\
p-n
\end{matrix}\right).\la{Gapp}
\ee
We can establish, from this expression, an important symmetry property of $G(s,s',p)$ valid when $p\neq 0,1$.
We have that  under the exchange $s'+2\leftrightarrow p$ while keeping  $s+s'$ and  $(p-s)$ fixed $G$ satisfies
\be 
G(s,s',p) &= \frac{(s'+2)(s'+1)}{p(p-1)} G(s+s'+2 - p, p-2, s'+2 ) .
\ee 
To evaluate \eqref{Gapp} there are four different cases to consider: 
$i) s'+2, s\geq p$, $ii) s\geq p \geq s'+2$,  $iii) s'+2 \geq p \geq s$, $iv) p\geq s, s'+2 $,
each of which leads to different summation ranges. 
In case $i)$ we have 
\be 
G(s,s',p)&:=\frac{(-1)^p(s'+2)!}{p! s'!}\sum_{n=0}^{p} (-)^{n}(s-n+1)
\left(\begin{matrix}
p\\
n
\end{matrix}\right)
\frac{\Gamma(s'+1+n)}{\Gamma(s'+3 -p +n)}\cr
&= \frac{(-1)^p(s'+2)!}{p!} 
(s+1 -\delta)\,{}_2\tilde{F}_1[-p,s'+1;s'+3 -p ;1]\,,
\ee
where ${}_2\tilde{F}_1$ is the regularized hypergeometric function given by 
$
{}_2\tilde{F}_1[a,b;c;z] := \frac{{}_2{F}_1[a,b;c;z]}{\Gamma(c)}$. In the last line above, $\delta=z\pa_z$ is a derivative operator,\footnote{ The notation  means $\delta {}_2\tilde{F}_1[a,b;c;1]:= (z\pa_z {}_2F_1[a,b;c;z])|_{z=1}$.}
that can be evaluated using {Gauss's summation formula}
\be\la{2F1G}
{}_2F_1[a,b,c;1]=\frac{\Gamma(c)\Gamma(c-a-b)}{\Gamma(c-a)\Gamma(c-b)}\,, \quad {{\rm Re}(c) > {\rm Re}(a + b)}\,,
\ee
and 
$\delta {}_2 F_1[a,b;c;1]= \frac{ab}{c-a-b-1} {}_2F_1[a,b;c;1]$.
When $a=-p$ is a negative integer we have
\be \la{2F1sum}
 {}_2\tilde{F}_1[-p,b;c;1]=\f{1}{\Gamma(b)} \sum_{n=0}^{p}(-1)^n \left(\begin{matrix}
p\\
n
\end{matrix}\right)
\frac{\Gamma(b+n)}{\Gamma(c +n)}.
\ee 
We thus find
\be
G(s,s',p)=
\f{(-1)^p(s+1 +p(s'+1))}{p!\Gamma(2 -p)}\,,
\ee
which is non-vanishing only for $p=0,1$.

 Similarly, in case $ii)$ we have that 
\be 
G(s,s',p)&:=\frac{(-1)^{s'}}{s'!}
\sum_{m=0}^{s'+2} (-)^{m}(s +s'+3-p -m)
\left(\begin{matrix}
s'+2\\
m
\end{matrix}\right)
\frac{\Gamma(m+p-1)}{\Gamma(m+p-s'-1 )}\\
&= \frac{(-1)^{s'}(p-2)!}{s'!} 
(s +s'+3-p -\delta)\,{}_2\tilde{F}_1[-(s'+2),p-1;p-s'-1 ;1] \cr
&=0\qquad {\rm for ~all}\quad  s\geq p \geq s'+2\,.
\ee
After an analogous analysis one finds that for both cases $iii)$ and $iv)$
\be 
G(s, s', p) =  \dfrac{(-1)^{p + s}\Gamma(3 + s + s')}{p (p - 1)\Gamma(p - 1 - s)\Gamma(1 + s)\Gamma(3 + s + s' - p)}. 
\ee
Putting everything together, the coefficient $G$ takes the form
\be 
G(s,s',p) &= \f{(-)^p ( s+1 + p (s'+1))}{p! \Gamma(2-p) } \quad {\rm{if}}\quad p \leq s, \la{G1}\\
G(s,s',p) &=  \f{(-)^{p+s}(s+s'+2)!}{\Gamma(p-s-1)(s+s'+2-p)! s!}
\f1{p(p-1)} \quad {\rm{if}}\quad p \geq s+1.\la{G2}
\ee

This result is suggestive of the split of the sums in \eqref{qq} as
\be
\sum_{p=0}^{s+s'+2} \sum_{n = {\rm max}[0,p-s'-2]}^{{\rm min}[p,s]} = \sum_{p=0}^{s} \sum_{n = {\rm max}[0,p-s'-2]}^{p}
+\sum_{p=s+1}^{s+s'+2} \sum_{n = {\rm max}[0,p-s'-2]}^{s}\,.
\ee
The first sum over $n$ is given in \eqref{G1} 
and it thus gives 
non-zero contribution only for $p=0,1$. The second sum over $n$ corresponds to the case \eqref{G2}. 
Finally, this allows us to write 
the charge bracket as\footnote{The second sum over $n$ vanishes for $p=s+1$. }
\be
\{q^2_s(z),  q^1_{s'}(z')\}
&=\frac{\kappa^2}{8}\bigg[ 
 (s+1)D_{z'}  q^1_{s'+s-1}(z')  \delta(z, z')
 -(s+s'+2) q^1_{s'+s-1}(z') D_{z} \delta(z, z')\cr
&+\sum_{p=s+2}^{s+s'+2}
\f{(-)^{p+s}(s+s'+2)!}{(p-s-2)!(s+s'+2-p)! s!}
\f1{p(p-1)}
\left( D_{z'}^{1-p}  q^1_{s'+s-1}(z')\right) D_{z}^{p} \delta(z, z')\bigg]\,.\cr
\ee

In order to compute the charge bracket at linear order we  need the antisymmetrize in $s,s'$ and $z,z'$, namely we need also the bracket
\be
\{q^1_s(z),  q^2_{s'}(z')\}&=-\{q^2_{s'}(z'),  q^1_{s}(z)\}\cr
&=-\frac{\kappa^2}{8}  \sum_{p=0}^{s+s'+2} \sum_{n = {\rm max}[0,p-s-2]}^{{\rm min}[p,s']} (-)^{p+n}(s'-n+1)
\left(\begin{matrix}
s+n\\
s
\end{matrix}\right)
\left(\begin{matrix}
s+2\\
p-n
\end{matrix}\right)\cr
&\times
\left( D_{z}^{1-p}  q^1_{s'+s-1}(z) D_{z'}^{p} \delta(z, z')\right)\,.
\ee
In this case the two sums can be split as
\be
\sum_{p=0}^{s+s'+2} \sum_{n = {\rm max}[0,p-s-2]}^{{\rm min}[p,s']} = 
\sum_{p=0}^{s'} \sum_{n = {\rm max}[0,p-s-2]}^{p}
+\sum_{p=s'+1}^{s+s'+2} \sum_{n = {\rm max}[0,p-s-2]}^{s'}\,.
\ee
The first sum over $n$ again gives non-zero contribution only for $p=0,1$. Antisymmetrizing \eqref{G1}, \eqref{G2}, we find
\be
\{q^1_s(z),  q^2_{s'}(z')\}
&=-\frac{\kappa^2}{8} \bigg[
(s'+1)D_{z}  q^1_{s'+s-1}(z)  \delta(z, z')
 -(s+s'+2) q^1_{s'+s-1}(z) D_{z'} \delta(z, z')\cr
 &+\sum_{p=s'+2}^{s+s'+2}
\f{(-)^{p+s'}(s+s'+2)!}{(p-s'-2)!(s+s'+2-p)! s'!}
\f1{p(p-1)}
\left( D_{z}^{1-p}  q^1_{s'+s-1}(z)\right) D_{z'}^{p} \delta(z, z')\bigg]\,.\cr
\ee
Combining the two brackets,
\be
&\{q^2_s(z),  q^1_{s'}(z')\}
+\{q^1_s(z),  q^2_{s'}(z')\}\cr
&=
\frac{\kappa^2}{8}\bigg[ 
\sum_{p=0}^{s+s'+2}
G(s,s',p)
\left( D_{z'}^{1-p}  q^1_{s'+s-1}(z')\right) D_{z}^{p} \delta(z, z')\cr
&-(s'+1)D_{z}  q^1_{s'+s-1}(z)  \delta(z, z')
 +(s+s'+2) q^1_{s'+s-1}(z) D_{z'} \delta(z, z')\cr
 &-\sum_{p=s'+2}^{s+s'+2}
\f{(-)^{p+s'}(s+s'+2)!}{(p-s'-2)!(s+s'+2-p)! s'!}
\f1{p(p-1)}
\left( D_{z}^{1-p}  q^1_{s'+s-1}(z)\right) D_{z'}^{p} \delta(z, z')\bigg]\,.
\la{abba}
\ee
Let us  rewrite
\be
\left( D_{z}^{1-p}  q^1_{s'+s-1}(z)\right) D_{z'}^{p} \delta(z, z')
&=D_{z'}^{p} \left[\left( D_{z}^{1-p}  q^1_{s'+s-1}(z)\right) \delta(z, z')\right]\cr
&=D_{z'}^{p} \left[\left( D_{z'}^{1-p}  q^1_{s'+s-1}(z')\right) \delta(z, z')\right]\cr
&=\sum_{m=0}^p \f{(-)^m p!}{m!(p-m)!}\left( D_{z'}^{1-m}  q^1_{s'+s-1}(z')\right) D_{z}^{m}\delta(z, z')\,,
\ee
so that the last term in \eqref{abba} becomes
\be
&\sum_{p=s'+2}^{s+s'+2}
\f{(-)^{p+s'}(s+s'+2)!}{(p-s'-2)!(s+s'+2-p)! s'!}
\f1{p(p-1)}
\left( D_{z}^{1-p}  q^1_{s'+s-1}(z)\right) D_{z'}^{p} \delta(z, z')\cr
&=\sum_{p=s'+2}^{s+s'+2}\sum_{m=0}^p
\f{(s+s'+2)!}{s'!m!} \f{(-)^{p+s'+m} (p-2)!}{(p-s'-2)!(s+s'+2-p)!(p-m)! }
\left( D_{z'}^{1-m}  q^1_{s'+s-1}(z')\right) D_{z}^{m}\delta(z, z')\cr
&=\sum_{m=0}^{s+s'+2}\sum_{p={\rm max}[m,s'+2]}^{s+s'+2}
\f{(-)^{s'+m}(s+s'+2)!}{s'!m!} \f{(-)^{p}( p-2)!}{(p-s'-2)!(s+s'+2-p)!(p-m)! }
\cr
&\times \left( D_{z'}^{1-m}  q^1_{s'+s-1}(z')\right) D_{z}^{m}\delta(z, z')\cr
&=\sum_{m=0}^{s+s'+2}\tilde G(s',s,m)\left( D_{z'}^{1-m}  q^1_{s'+s-1}(z')\right) D_{z}^{m}\delta(z, z')\,,
\ee
where we have introduced the coefficient
\be 
\tilde{G}(s',s,m)&:=\sum_{p={\rm max}[m,2]}^{s+s'+2} (-)^m  \left(\begin{matrix} p\\ m\end{matrix}\right)G(s',s,p)\cr
&= \sum_{p={\rm max}[m,s'+2]}^{s+s'+2}\f{(-)^m p!}{m!(p-m)!} \f{(-)^{p+s'}(s+s'+2)!}{(p-s'-2)!(s+s'+2-p)! s'!} \f1{p(p-1)} \cr
&= \f{(-)^m (s+s'+2)!}{m!s'!} \sum_{p={\rm max}[m,s'+2]}^{s+s'+2} 
\f{ }{} \f{(-)^{p+s'}(p-2)! }{(p-s'-2)!(s+s'+2-p)! (p-m)! } \cr
&=
\left\{\begin{array}{l}
\f{(-)^{m} (s+s'+2)!}{m!s!} {}_2\tilde F_1[-s,s'+1;s'+3-m;1],\quad {\rm{if}}\quad m\leq s'+2
\vspace*{2mm}\\
\f{(-)^{s'}(s+s'+2)! (m-2)!}{m! s'!(s+s'+2-m)!} {}_2\tilde F_1[-(s+s'+2-m),m-1;m-s'-1;1], \quad{\rm{if}}\quad m\geq s'+2\,.
\end{array}
\right.\cr
\ee 
The evaluation of the regularized hypergeometric functions  for $m\leq s'+2$ and $m\geq s'+2$    gives respectively
\be 
&{}_2\tilde{F}_1[-s,s'+1;s'+3-m;1]  = \frac{\Gamma(s+2-m)}{\Gamma(2-m)\Gamma(s+s'+3-m)}\,,\cr
&{}_2\tilde{F}_1[-(s+s'+2-m),m-1;m-s'-1;1] =\frac{\Gamma(s+2-m)}{\Gamma(-s') \Gamma(s+1)},
\ee 
where we used that ${}_2\tilde F_1[a,b,c;1]=\frac{\Gamma(c-a-b)}{\Gamma(c-a)\Gamma(c-b)}$.
This means that $\tilde{G}(s',s,m)=0$ if $2\leq m \leq s+1$.\footnote{Irrespective of whether  $m\geq s'+2$ or $m\leq s'+2$.} It also implies
that 
\be
\tilde{G}(s',s,m)=\frac{(-)^{s-m}}{m(m-1)} 
\f{(s+s'+2)!}{(s+s'+2-m)!s!(m-s-2)!}\,,
\ee 
if $m\geq s+2$, and 
\be 
\tilde{G}(s',s,0)=(s+1),
\qquad
\tilde{G}(s',s,1)= -(s+s'+2).
\ee 
In other words, we can establish the identity 
\be\la{OMNEG}
\tilde{G}(s',s,m)= G(s,s',m) \qquad {\rm for~all}\quad m. 
\ee

Therefore, the bracket \eqref{abba} finally becomes
\be
&\{q^2_s(z),  q^1_{s'}(z')\}
+\{q^1_s(z),  q^2_{s'}(z')\}\cr
&=\frac{\kappa^2}{8}\bigg[ 
\sum_{p=0}^{s+s'+2}
\left(G(s,s',p)-\tilde G(s',s,p)\right)
\left( D_{z'}^{1-p}  q^1_{s'+s-1}(z')\right) D_{z}^{p} \delta(z, z')\cr
&-(s'+1)D_{z}  q^1_{s'+s-1}(z)  \delta(z, z')
 +(s+s'+2) q^1_{s'+s-1}(z) D_{z'} \delta(z, z')
 \bigg]\cr
 &=\frac{\kappa^2}{8}\left[ 
-\left(s'+1\right)  q^1_{s'+s-1}(z') D_{z} \delta(z, z')
+ (s+1) 
q^1_{s'+s-1}(z) D_{z'} \delta(z, z')
\right]\,,
\ee
where we have used \eqref{OMNEG} in the last passage. This concludes the proof of \eqref{WTF}.

We conclude this appendix with a proof for the relation \eqref{DD}. We have
\be
\sum_{\ell = n}^{s} \frac{(\ell + 1)! (\Delta - \ell)_{s-\ell}}{(\ell - n)! (s-\ell)!} 
&=\sum_{\ell = n}^{s} \frac{(\ell + 1)! \Gamma(\Delta - \ell+1) }{(\ell - n)! (s-\ell)! \Gamma(\Delta - s+1) }\cr
&=\sum_{p = 0}^{s-n} \frac{(p+n + 1)! \Gamma(\Delta - p-n+1) }{p! (s-p-n)! \Gamma(\Delta - s+1) }\cr
&=\sum_{p = 0}^{s-n} \f{(-)^{p+n+s}}{(s-n)!}
 \left(\begin{matrix} s-n\\ p\end{matrix}\right)\f{\Gamma(p+n+2) \Gamma(s-\Delta)}{\Gamma(p+n-\Delta)}
 \cr
 &=\f{(-)^{n+s}(n+1)!}{(s-n)!}  \Gamma(s-\Delta)  {}_2\tilde{F}_1[n-s,n+2;n-\Delta;1]\cr
 &=\f{(-)^{n+s}(n+1)!}{(s-n)!} \f{ \Gamma(s-\Delta-n-2 )}{\Gamma(-2-\Delta)}\cr
 &=\f{(n+1)!}{(s-n)!} \f{ \Gamma(\Delta+3 )}{\Gamma(\Delta+3-s+n)}\cr
&=  \frac{(n+1)!}{(s-n)!} (\Delta + 2)_{s - n}\,,
\ee
where we used \eqref{2F1G}, \eqref{2F1sum}, and twice  the property
\be
\Gamma(\alpha-n)=(-)^{n-1}\f{\Gamma(-\alpha)\Gamma(1+\alpha)}{\Gamma(n+1-\alpha)}\,,\qquad n\in\Z\,.
\ee

\section{Normalization}\label{App:Norm}

The (outgoing) conformal primary gravitons are typically defined as Mellin transforms of asymptotic particle states, 
\be 
O_{\Delta}^{\pm}(z, \bz) = \int_0^{\infty} d\omega \omega^{\Delta - 1} \langle \omega, z, \bz | = \int_0^{\infty} d\omega \omega^{\Delta - 1} \langle 0| a^{\rm out}_{\pm}(\omega, z, \bz). 
\ee
Then using 
\be 
\widetilde{C}(\omega, z) = \int du e^{i\omega u} C(u, z) = \frac{i\kappa}{4\pi} \left[a_-^{{\rm out} \dagger}(\omega \hat{x})\theta(-\omega) - a_+^{\rm out}(\omega\hat{x})\theta(\omega) \right],
\ee
we find that
\be 
\label{phcsg}
\begin{split}
O_{\Delta}^{+}(z, \bz) &= -\frac{4\pi}{i\kappa} \int_0^{\infty} d\omega \omega^{\Delta - 1} \widetilde{C}(\omega, z) = -\frac{4\pi}{i\kappa}  \int_0^{\infty} d\omega \omega^{\Delta - 1} \int_{-\infty}^{\infty} du e^{i\omega (u+i\epsilon)} C(u,z) \\
&=  -i^{\Delta} \frac{4\pi}{i\kappa}  \Gamma(\Delta) \int_{-\infty}^{+\infty} du  (u{+i\epsilon})^{-\Delta}  C(u,z)
= i^{\Delta} \frac{8\pi}{i\kappa} G_{\Delta}^+(z). 
\end{split}
\ee
where 
\be 
G_{\Delta}^+(z)= -\frac{\Gamma(\Delta)}{2} \int_{-\infty}^{+\infty} du  (u+i\epsilon)^{-\Delta}  C(u,z)
\ee 
Similarly, 
$O_{\Delta}^{-}(z, \bz)=i^{\Delta} \frac{8\pi}{i\kappa} G_{\Delta}^{-}(z, \bz)$, where
\be 
\label{nhcsg}
\begin{split}
G_{\Delta}^{-}(z, \bz) &=  - \frac{  \Gamma(\Delta)}{2} \int_{-\infty}^{+\infty} du  (u{+}i\epsilon)^{-\Delta}  \bar{C}(u,z). 
\end{split}
\ee

Pseudo-differential calculus identities can be easily proven in a conformal primary basis, where for instance 
$C\to \pa_u C$ corresponds to
$G_{\Delta}\to G_{\Delta+1}$, while 
$C\to u C$ corresponds to
$G_{\Delta}\to G_{\Delta-1}$.
For example, this allows us to have a simple proof of the identity \eqref{pip}
\be 
&\p_{u}^{1-\ell} \left(C(u) \frac{u^{s-\ell}}{(s-\ell)!} \right)\rightarrow 
\Gamma(\Delta) \int \rd u u^{-\Delta}  \p_{u}^{1-\ell} \left(C(u) \frac{u^{s-\ell}}{(s-\ell)!} \right) \cr  
&=  \Gamma(\Delta -\ell+1) \int \rd u u^{-\Delta +\ell -1}  \left(C(u) \frac{u^{s-\ell}}{(s-\ell)!} \right) \cr
&=  \frac{\Gamma(\Delta -\ell +1)}{(s-\ell)! \Gamma(\Delta -s +1)} G^+_{\Delta +1-s} 
=\frac{(\Delta - \ell)_{s-\ell} }{(s-\ell)!}  G^+_{\Delta +1-s}.
\ee

\section{$w$-current }\la{App:wcurr}

In this appendix we give the proof of \eqref{Lightt}.
We start from the expansion
\be
 \epsilon \,G^{{-}}_{1-s +\epsilon}(z,\bz) = z^{s+1 -\epsilon } \sum_{m=0}^\infty \frac{ N_s^{m -s - 1}(\bz)}{z^m}\,,
\ee
which implies that 
\be
&(-1)^{(s+3)} \Gamma(s+3 ) {\rm \bf L}\left[G_{ 1-s + \epsilon}   \right](z,\bz) \cr
&=\frac{1}{\epsilon}\sum_{n= -(s+1)}^{\infty} N_{s}^n(\bz)
\int_{\mathbb{R}} \frac{d {w}}{2\pi i} \frac{(-1)^{(s+3)} \Gamma(s+3)}{(z - {w})^{s+3 - \epsilon} w^{(n +\epsilon)}} \cr
&= \sum_{n= -(s+1)}^{\infty} \frac{N_s^n}{z^{(s+2+n)}} \frac{(-1)^{s} \Gamma(s+2+n)}{\epsilon \Gamma(n+\epsilon)}\cr
&\sim \underbrace{\sum_{n= 0 }^{(s+1)}\frac{(-1)^{(n+s)} N_s^{-n}}{z^{(s+2-n)}} n! (s+1-n )!}_{W_s}
+ \frac1\epsilon \underbrace{\sum_{n= 0 }^{\infty}\frac{(-1)^{s} N_s^n}{z^{(s+2+n)}} \frac{(s+1+n )!}{(n-1)!}}_{q_s^1}.
\ee 
In the second equality we have used the relation \cite{Himwich:2021dau}
\be 
\int_{\mathbb{R}} \frac{d {w}}{2\pi i} \frac{1}{(1 - {w})^{a} w^{b}}
= -\frac{\Gamma(a+b-1)}{\Gamma(a)\Gamma(b)},
\ee
valid after analytic continuation from the domain $\mathrm{Re}(a)<1$, $\mathrm{Re}(b)<1$ and $\mathrm{Re}(a+b)>1$.
Note that in the last term we have neglected terms that arises from $\pa_\epsilon N_{s+\epsilon}|_{\epsilon=0}$.

\bibliographystyle{bib-style2.bst}
\bibliography{biblio-w.bib}

\end{document}